# Keyword Decisions in Sponsored Search Advertising: A Literature Review and Research Agenda

Yanwu Yang[1] and Huiran Li[2]

[1]School of Management, Huazhong University of Science and Technology, Wuhan 430074, China

[2]School of Business Administration and Customs Affair, Shanghai Customs College, Shanghai 201204, China

{yangyanwu.isec, lihuiran.isec}@gmail.com

**Abstract.** In sponsored search advertising (SSA), keywords serve as the basic unit of business model, linking three stakeholders: consumers, advertisers and search engines. This paper presents an overarching framework for keyword decisions that highlights the touchpoints in search advertising management, including four levels of keyword decisions, i.e., domain-specific keyword pool generation, keyword targeting, keyword assignment and grouping, and keyword adjustment. Using this framework, we review the state-of-the-art research literature on keyword decisions with respect to techniques, input features and evaluation metrics. Finally, we discuss evolving issues and identify potential gaps that exist in the literature and outline novel research perspectives for future exploration.

# 1. Introduction

Sponsored search advertising (SSA) has become one of the most successful business models of online advertising. Millions of advertisers spent a large amount of advertising budgets in SSA to promote their products and services (Yang et al., 2018). According to a recent IAB report (Interactive Advertising Bureau, 2022), in the United States alone, the annual internet advertising revenue of 2021 reached $189.3 billion where SSA accounts for around 41.4% of that pie.

In SSA, firms need to choose suitable keywords to describe their products or services efficiently, and organize these keywords following certain advertising structures (e.g., account, ad-campaign, and ad-group) defined by major search engines. Once a user submits a query to a search engine which is related to one or several of these keywords, it triggers an auction process that determines which advertisements and their rankings to be displayed on search engine result pages (SERPs), together with a set of organic search results. In the SSA ecosystem, keywords are the unique carriers connecting advertisers, potential consumers, and search engines. Moreover, keywords are the basic units for advertisers to conduct online market research, design and evaluate marketing strategies. Keywords play a crucially important role in business competition for companies in online platforms. In practice, advertisers have to make various keyword decisions throughout the entire lifecycle of SSA campaigns (Yang et al., 2019). Therefore, it becomes a critical issue for search advertisers to make a series of effective keyword decisions in SSA.

Since the advent of SSA, keyword decisions have increasingly attracted research interests from both academia and industries. As far as we knew, on one hand, it is apparent that there are no commonly agreed definitions for related concepts identified in the extant literature on keyword decisions; on the other hand, prior research on keyword decisions has been conducted either separately on an individual keyword decision or without consideration of search advertising structures. There is a need for developing an integrated review of the state-of-the-art knowledge about keyword decisions in SSA. Our objectives for this paper are to examine what has been done in the literature on keyword decisions, uncover the potential gaps, and figure out novel research perspectives for future exploration.

This review complements recent review articles on online advertising and advertising selection. Ha (2008) conducted a review of online advertising published in major advertising journals from 1996 to 2007, which focuses on analyzing conceptual foundations, theories, and state-of-the-art practices of online advertising. Shatnawi & Mohamed (2012) presented an overview of online advertising selection, which focuses on investigating existing approaches, comparing and classifying these approaches. Our review focuses on keyword decisions in SSA from the system perspective, by taking into account search advertising structures and the entire lifecycle of advertising campaigns.

The contribution of our review can be summarized in the following ways. First, to the best of our knowledge, this review is the first effort focusing on keyword decisions in SSA, which has not been systematically explored before. We present an overarching framework for keyword decisions based on practical decision scenarios throughout the entire lifecycle of SSA campaigns, and conduct a systematic review of the state-of-the-art literature with respect to techniques, input features and evaluation metrics. Second, we find that a lot of practical issues remain unaddressed in this field, although plenty of research efforts have been involved in the last two decades. In particular, few research efforts reported on practical keyword decisions such as keyword targeting, keyword assignment and grouping, and keyword adjustment. In addition, our review will provide foundations for continuing studies on keyword decisions in SSA and other advertising forms.

## 2. Survey Scope and Structure

### 2.1 Survey Scope

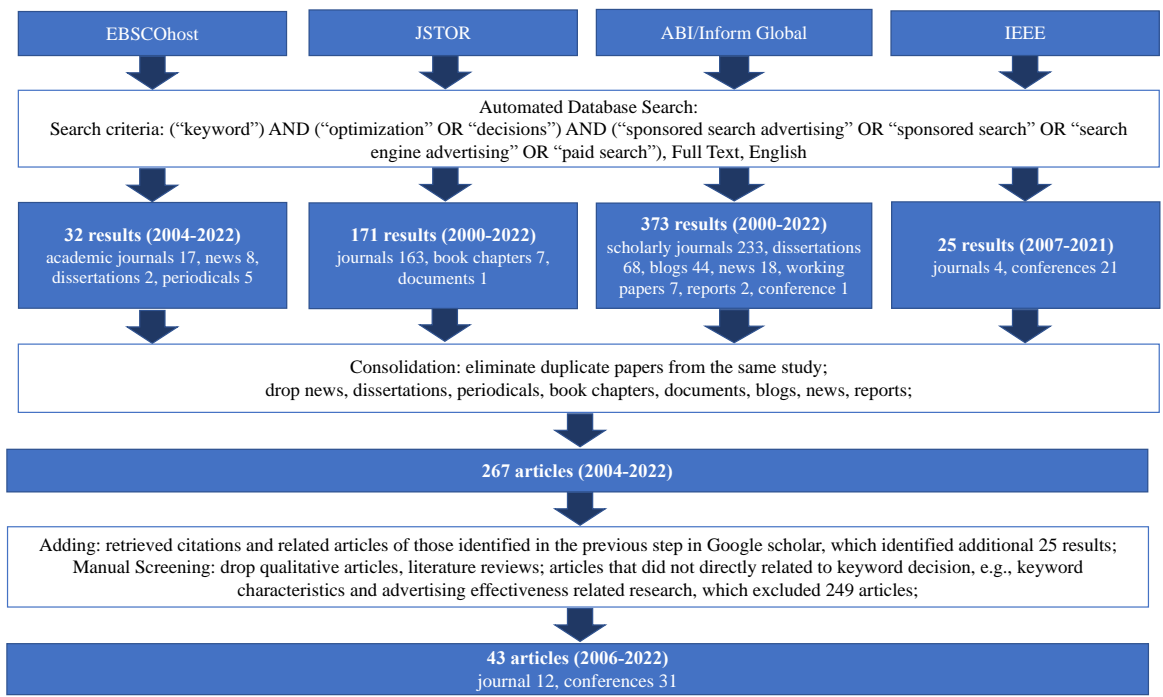

Figure 1. Study Search and Selection

Due to the interdisciplinary nature of keyword decisions topics, research articles covered in this review were published in major IT-oriented (e.g., computer science, artificial intelligence and information retrieval) and/or business-oriented (e.g., management information systems, advertising and marketing) journals and conferences. Our review includes articles retrieved mainly from four academic databases: EBSCOhost, JSTOR, ABI-Inform and IEEE using full text search of ("keyword") AND ("optimization" OR "decisions") AND ("sponsored search advertising" OR "sponsored search" OR "search engine advertising" OR "paid search"). We restricted our selected studies in English. The search resulted in 32 results from EBSCOhost, 171 results from JSTOR, 373 results from ABI/Inform Global and 25 results from IEEE. We manually screened the results to eliminate duplicate articles, news, dissertations, periodicals, book chapters, documents, blogs, news and reports. This process led to 267 articles. Moreover, we expanded the literature by retrieving citations of the articles obtained in the previous step in Google scholar, which yielded additional 25 results. Furthermore, we dropped qualitative articles, literature reviews and empirical articles on keyword research, by going through title, abstract, full-text of each article, which excluded 249 articles. Finally, this literature search resulted in a selection of 43 research publications, including 12 peer-reviewed journal articles and 31 conference articles, covering the period from 2006 to 2022. Our search process is illustrated in Fig. 1.

**2.2 Survey Structure**

In SSA, a general advertising structure employed by major search engines (e.g., Google, Bing) can be described as: under a SSA account of an advertiser, there are one or several campaigns that are run simultaneously in order to fulfill a promotional goal; in the meanwhile, one or several ad-groups comprise a campaign, and each ad-group consists of one or more ad-copies and a set of keywords. Fig. 2 presents an illustration of the search advertising structure. In this sense, SSA is essentially distinct from traditional advertising (e.g., print ads and TV ads) due to its hierarchical advertising structure. Thus, advertising decisions in SSA are essentially structured, rather than flatted as in traditional advertising (Yang et al., 2012, 2019).

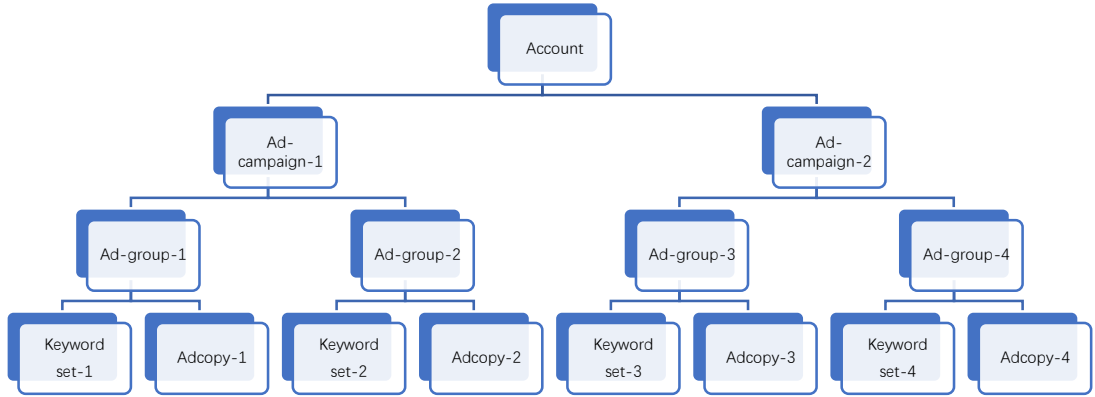

Figure 2. An Illustration of Sponsored Search Advertising Structure

Throughout the entire lifecycle of SSA campaigns, advertisers have to make a series of keyword related decisions at different levels, namely keyword generation at the domain level, keyword targeting at the market level, keyword assignment and grouping at the campaign and ad-group level, and dynamical keyword adjustment, forming a closed-loop decision cycle (Yang et al., 2019). In this review, we organize the extant literature on keyword decisions within a research framework, as presented in Fig. 3.

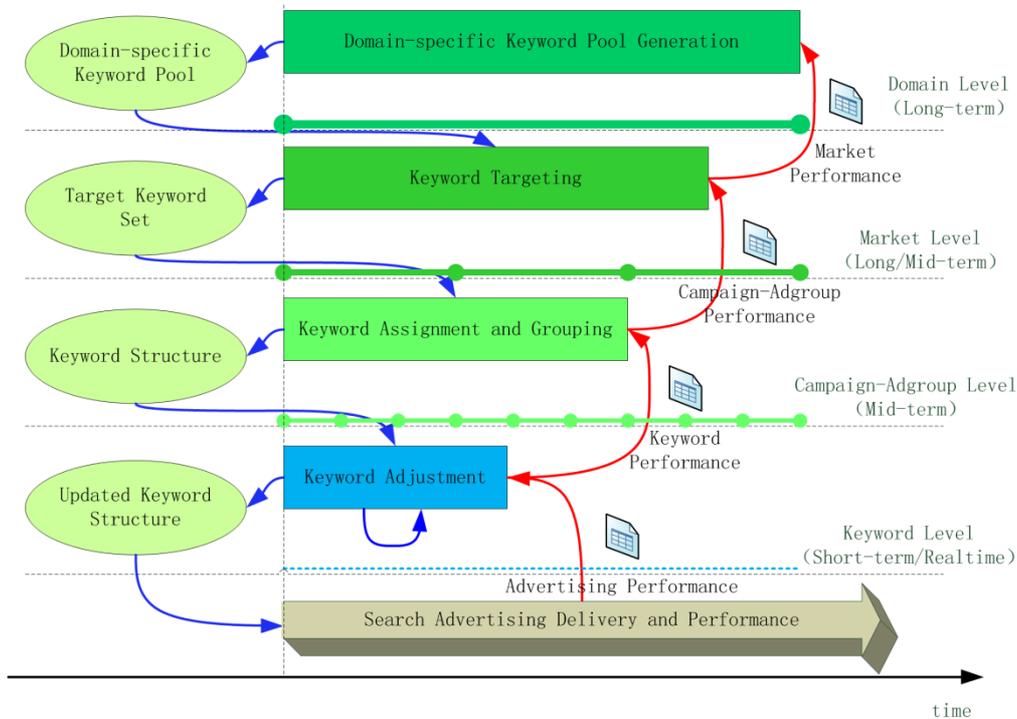

Figure 3. The Framework of Our Literature Review on Keyword Decisions

Section 3 centers on keyword generation, which is also known as the domain-level keyword optimization, aiming to generate a domain-specific keyword pool. In this section, after defining the

keyword generation problem, we go over techniques, features and evaluation metrics used in the literature on keyword generation.

Section 4 discusses keyword targeting, which is also known as the market-level keyword optimization, aiming to obtain a more targeted set of keywords from the domain-specific keyword pool and choose appropriate match types for selected keywords in order to reach the right population of potential consumers. In this section, we first define the keyword targeting problem, and then review techniques, features and evaluation metrics for keyword selection and keyword match.

Section 5 focuses on keyword assignment and grouping, which is also known as the campaign and ad-group level keyword optimization, aiming to yield an effective keyword structure by following the advertising structure of SSA. In this section, we first define the problem of keyword assignment and grouping, and then introduce techniques, features and evaluation metrics used in this area.

Section 6 concerns keyword adjustment, i.e., advertisers have to dynamically adjust their keyword structures and associated strategies according to the realtime advertising performance of SSA campaigns. In this section, we first define the keyword adjustment problem in more detail, and then we discuss the status of this area.

Section 7 summarizes the state-of-the-art keyword decisions, uncovers the gaps that exist in the literature, and suggests promising research perspectives for future exploration. Finally, we conclude our review in Section 8.

Notations used in problem definitions for keyword decisions are presented in Table 1.

Table 1. Notations in problem definitions for keyword decisions

| Terms | Definition |
|---|---|
| $f^{(GNT)}$ | The keyword generation function |
| $S^{(GNT)}$ | Information source for keyword generation |
| $K^{(GNT)}$ | A set of generated keywords |
| $k_i$ | The $i$-th keyword in a set |
| $n^{(GNT)}$ | The number of generated keywords |
| $f^{(TGT)}$ | The keyword targeting function |
| $n^{(TGT)}$ | The number of targeting keywords |
| $K^{(TGT)}$ | A set of selected keywords |
| $x_{i,\bar{m}}^{(TGT)}$ | A binary decision variable of keyword targeting, indicating whether the $i$-th keyword in match type $\bar{m}$ is selected |
| $\bar{m}$ | Keyword match type |
| $f^{(ASM)}$ | The keyword assignment function |
| $K_j^{(ASM)}$ | A set of keywords assigned to the $j$-th campaign |
| $x_{i,j}^{(ASM)}$ | A binary decision variable of keyword assignment, indicating whether the $i$-th keyword is assigned to the $j$-th campaign |
| $n_{campaign}$ | The number of ad campaigns |
| $f^{(GRP)}$ | The keyword grouping function |
| $K_{j,l}^{(GRP)}$ | A set of keywords for the $l$-th ad-group of the $j$-th campaign |
| $x_{i,j,l}^{(GRP)}$ | A binary decision variable of keyword grouping, indicating whether the $i$-th keyword is grouped into the $l$-th ad-group of the $j$-th campaign |
| $n_{group}$ | The number of ad-groups |
| $f^{(ADJ)}$ | The keyword adjustment function |
| $x_{i,j,l,t}^{(ADJ)}$ | A binary decision vector of keyword adjustment, indicating whether the $i$-th keyword is grouped into the $l$-th ad-group of the $j$-th campaign at time $t$ |
| $K_{j,l,t}^{(ADJ)}$ | A set of keywords grouped into the $l$-th ad-group of the $j$-th campaign at time $t$ |

## 3. Domain-Specific Keyword Pool Generation

### 3.1 Problem Description

For advertisers in an application domain, they need to build a pool of relevant keywords over which they can conduct marketing research and from which a more accurate set of keywords can be determined for their search advertising campaigns. This step is called keyword generation, and its output is the domain-specific keyword pool.

Formally, the keyword generation problem can be defined as follows. Let $S^{(GNT)}$ denote information sources for keyword generation such as a corpus of Web pages, query logs, search results, advertising database, domain-specific semantics and concept hierarchy, the keyword generation process can be given as

$$f^{(GNT)}: S^{(GNT)} \to K^{(GNT)} = \{k_1, \dots, k_i, \dots, k_{n^{(GNT)}}\}, i \in \{1, 2, \dots, n^{(GNT)}\}, \quad (1)$$

where $f^{(GNT)}$ denotes the keyword generation function, $K^{(GNT)}$ denotes a set of generated keywords and $k_i$ denotes the $i$-th keyword in the set.

Generally, keyword pool generation aims to obtain a set of keywords that represents the domain-specific knowledge and information of the targeted market as comprehensive and relevant as possible (Nie et al., 2019), starting with websites (or Web pages) or a set of seed keywords provided by advertisers. In this sense, keyword generation is also known as keyword expansion in that it expands from one or several seed keywords. The generated set of keywords can be viewed as an extended business description for advertisers.

A plethora of academic efforts have been devoted to keyword generation, in order to help advertisers reach potential consumers (e.g., Yih et al., 2006; Thomaidou and Vazirgiannis, 2011; Nie et al., 2019; Scholz et al., 2019; Yang et al., 2019). Note that the keyword generation problem of SSA is essentially similar to that of online contextual advertising in that both are intended to obtain a set of keywords of interest to consumers. Thereby, prior research in this direction did not make a distinction between them. Hence, we follow such a convention in this paper by reviewing keyword generation for the two advertising contexts within a same framework. A related research stream to keyword generation is query expansion in the field of information retrieval. For an extensive review on query expansion, refer to see Carpineto & Romano (2012) and Azad & Deepak (2019).

In the following, we present major techniques, and then discuss input features and evaluation metrics used in the literature.

**3.2 Techniques for Keyword Generation**

State-of-the-art keyword generation techniques reported in the SSA literature can be categorized into five groups: (1) co-occurrence statistics, (2) similarity measures, (3) multivariate models, (4) linguistics processing models, and (5) machine learning models. Table 2 presents categories of state-of-the-art keyword generation techniques and the distribution of selected studies in this section. In the following, we will go over the specific keyword generation techniques in each category, and analyze their advantages and disadvantages.

Table 2. Techniques for Keyword Generation

| Category | Approach | Reference | Sources |
|---|---|---|---|
| Co-occurrence statistics | / | Zhou et al. (2007) | Websites and Web pages |
| Similarity measures | The Jaccard similarity | Joshi & Motwani (2006) | Search result snippets |
| | | Mirizzi et al. (2010) | Semantics and concept hierarchy, search result snippets |
| | | Chen et al. (2008) | Semantics and concept hierarchy |
| | The cosine similarity | Abhishek & Hosanagar (2007) | Websites and Web pages, search result snippets |
| | | Chang et al. (2009); Sarmento et al. (2009) | Advertising databases |
| | | Thomaidou & Vazirgiannis (2011) | Search result snippets, websites and Web pages |

| | Semantic similarity between terms | Amiri et al. (2008) | Semantics and concept hierarchy, search result snippets, query logs |
|---|---|---|---|
| Multivariate models | Logistic regression | Yih et al. (2006); Wu &Bolivar (2008); Lee et al. (2009) | Websites and Web pages |
| | | Berlt et al. (2011) | Advertising databases |
| | | Bartz et al. (2006) | Query logs, advertising databases |
| | Collaborative filtering | Bartz et al. (2006) | Query logs, advertising databases |
| | Topic-sensitive PageRank | Zhang et al. (2012a) | Semantics and concept hierarchy, websites and Web pages |
| | Latent Dirichlet allocation | Qiao et al. (2017) | Query logs |
| | | Welch et al. (2010) | Semantics and concept hierarchy, search result snippets |
| | Hierarchical Bayesian | Nie et al. (2019) | Semantics and concept hierarchy |
| Linguistics processing | Translation model | Ravi et al. (2010) | Advertising databases, websites and Web pages, query logs |
| | Word sense disambiguation | Scaiano &Inkpen (2011) | Semantics and concept hierarchy |
| | The relevance-based language model | Jadidinejad &Mahmoudi (2014) | Semantics and concept hierarchy |
| | Heuristics-based method | Scholz et al. (2019) | Query logs |
| Machine learning | Random walk | Fuxman et al. (2008) | Query logs |
| | Decision tree | GM et al. (2011) | Websites and Web pages |
| | Sequential pattern mining | Li et al. (2007) | Websites and Web pages |
| | Active learning | Wu et al. (2009) | Search result snippets |
| | Bayesian online learning | Schwaighofer et al. (2009) | Advertising databases |
| | Multi-step semantic transfer analysis | Zhang &Qiao (2018); Zhang et al. (2021) | Semantics and concept hierarchy, query logs |
| | Sequence-to-sequence learning | Zhou et al. (2019) | Query logs |

### 3.2.1 Co-occurrence Statistics

Co-occurrence statistics count paired terms within a collection such as Web pages, taking Web advertising as an information retrieval problem. Notations used in co-occurrence statistics for keyword generation are presented in Table 3a.

Table 3a. Notations used in co-occurrence statistics for keyword generation

| Terms | Definition |
|---|---|
| $K$, $K'$ | A finite set of keywords/terms |
| $k$, $k'$ | A keyword/term |
| $freq(k_i, k_j)$ | The co-occurrence frequency of $k_i$ and $k_j$ |
| $K(tf)$ | The top frequent keywords/terms |

| $\bar{k}$ | The current keyword/term |
|---|---|
| $ratio(k)$ | The ratio of the sum of the total number of terms in sentences including $k$ to the total number of terms in the document |
| $n_{\bar{k}}$ | The total number of keywords/terms in sentences including $\bar{k}$ |

In keyword generation, the word co-occurrence matrix is constructed to get weighted keywords. The general setting of co-occurrence statistics is as follows. Given two finite sets of keywords $K$ and $K'$, $k_i \in K$, $k_j \in K'$, let $K \times K'$ denote the Cartesian product of $K$ and $K'$, the co-occurrence statistics of a keyword pair $(k_i, k_j)$ is given by $freq(k_i, k_j) = |\{(k, k') \in K \times K' | k = k_i, k' = k_j\}|$, which measures the frequency of co-occurrence of $k_i$ and $k_j$.

Zhou et al. (2007) proposed a keyword extraction model where Web pages and advertisements are represented in a same data structure to support a retrieval process. In their model, nouns and verbs are extracted from the text of Web pages and the co-occurrence frequency between each term and top frequent terms is counted. Specifically, the term weight is calculated as

$$\text{Chi} - \text{Square}(\bar{k}) = \sum_{k \in K(tf)} \frac{(freq(\bar{k},k) - n_{\bar{k}} ratio(k))^2}{n_{\bar{k}} ratio(k)} - \max_{k \in K(tf)} \frac{(freq(\bar{k},k) - n_{\bar{k}} ratio(k))^2}{n_{\bar{k}} ratio(k)}, \quad (2)$$

where $K(tf)$ is the top frequent terms; $k$ and $\bar{k}$ are a term in $K(tf)$ and the current term, respectively; $freq(\bar{k}, k)$ is the co-occurrence frequency of $k$ and $\bar{k}$; $n_{\bar{k}}$ is the total number of terms in sentences including $\bar{k}$; $ratio(k)$ is the ratio of the sum of the total number of terms in sentences including $k$ to the total number of terms in the document.

A higher $\text{Chi} - \text{square}(\bar{k})$ indicates that the current term (i.e., $\bar{k}$) is more important in representing the document. Generally, co-occurrence statistics is used as a baseline in the research on keyword generation.

**3.2.2 Similarity Measures**

In cases where two relevant keywords do not occur together, co-occurrence statistics may fail to generate right keywords. Moreover, a keyword may have more than one meaning (Chen et al., 2008), which makes it difficult to filter out generated keywords. Thus, similarity measures are developed to explore the characteristic of keywords. Notations used in similarity measures for keyword generation are presented in Table 3b.

Table 3b. Notations used in similarity measures for keyword generation

| Terms | Definition |
|---|---|
| $C[k_i]$ | A characteristic keyword vector |
| $W$ | A feature weighting function |

| $sim(k_i, k_j)$ | The similarity between $k_i$ and $k_j$ |
| --- | --- |
| $doc$ | A document |
| $DOC$ | A set of documents $DOC = \{doc_1, doc_2, \ldots, doc_n\}$ |
| $quality(k', k)$ | The quality of suggesting keyword $k'$ for query $k$ |
| $\boldsymbol{v}_i$ | The TFIDF keyword vector for $doc_i$ |
| $TF$ | The term frequency |
| $IDF$ | The inverse document frequency |
| $PMI(k)$ | A point-wise mutual information feature vector for keyword $k$ |
| $pmi_{k_i}$ | The pointwise mutual information between $k_i$ and a given keyword |

Specifically, each candidate keyword is represented as a characteristic keyword vector $C[k_i]$. Similarity between each pair of keywords is given as

$$sim(k_i, k_j) = sim\_metric(W(C[k_i]), W(C[k_j])), \quad (3)$$

where $sim\_metric$ is a similarity measure, and one of the most popular possible instantiations of $sim\_metric$ is the cosine; $W$ is the feature weighting function. Term frequency-inverse document frequency (TFIDF) is commonly used as a weighting statistic, which is calculated as $TFIDF(k, doc, DOC) = TF(k, doc) \cdot IDF(k, DOC)$, where term frequency (TF) is the relative frequency of keyword $k$ within document $doc$, i.e., $TF(k, doc) = \frac{freq_{k,doc}}{\sum_{k' \in doc} freq_{k',doc}}$, and the inverse document frequency (IDF) is the logarithmically scaled inverse fraction of $DOC$ that contains $k$, i.e., $IDF(k, DOC) = -log \frac{|\{doc \in DOC : k \in doc\}|}{|DOC|}$.

A similarity graph can be constructed on the basis of the similarity between each pair of keywords, where nodes are keywords and the edges indicate similarities between keywords. Through traversing the similarity graph, a set of relevant but cheaper keywords can be generated (Joshi & Motwani, 2006; Abhishek & Hosanagar, 2007; Amiri et al., 2008; Thomaidou & Vazirgiannis, 2011).

In similarity-based keyword generation, one of common information sources is search result snippets. Given starting seeds, each keyword $k$ is submitted as a query to a search engine to retrieve a set of characteristic documents $DOC = \{doc_1, doc_2, \ldots, doc_n\}$, which is used to create a context vector for the input keyword and extract relevant keywords. For example, Joshi & Motwani (2006) characterized each input keyword using its text-snippets (i.e., words before and after the input keyword) document from the top 50 search-hits. The relevance of keyword $k'$ to $k$ is measured by the frequency of $k'$ observed in the characteristic document of keyword $k$; then the directed relevance between keywords was used to construct a directed graph (i.e., TermsNet). TermsNet suggests keywords through ranking their qualities. Specifically, the quality of $k'$ for $k$ is defined as $quality(k', k) = $ $freq(k',k)/(1 + \log(1 + \sum freq(k',i)))$, where each $i$ is an outneighbor of $k'$, which helps identify relevant yet nonobvious terms and their semantic associations. Abhishek & Hosanagar (2007) built a mathematical formulation which measures semantic similarity between keywords using a Web-based kernel function. They scraped advertisers' webpages, extracted and added keywords with high TFIDF into the initial dictionary, then created the context vector for each keyword. Given the TFIDF keyword vector $\boldsymbol{\mathcal{V}}_i$ for document $doc_i$, the context vector is the $L_2$ normalized centroid of $\frac{1}{n}\sum_{i=1}^{n}(\boldsymbol{\mathcal{V}}_i/\|\boldsymbol{\mathcal{V}}_i\|_2)$. The semantic similarity kernel function is defined as the inner product of context vectors for two keywords. Through extracting keywords from the landing page as initial seeds, Thomaidou & Vazirgiannis (2011) entered the extracted seeds into a search engine and parsed the snippets and titles of search results to construct a characteristic vector for these keywords. Most occurrences inside the resulted documents are kept as the most relevant keywords for the seed query. The proposed method can generate keywords that do not explicitly show on the landing page.

Similarity-based methods by exploring sources such as search result snippets can help increase the precision of keyword generation and catch the trend of consumer behaviors. However, when constructing the characteristic vectors of keywords, most similarity-based methods with search result snippets favor the keywords with high co-occurrence frequency. This might lead an expensive cost for the generated keywords, because keywords with high co-occurrence frequency are typically popular. Moreover, keywords generated by exploring the top-hit search result snippets are expensive as well (Thomaidou & Vazirgiannis, 2011). Thus, similarity-based keyword generation with search result snippets may place advertisers into a highly competitive environment, thus can't guarantee profit maximization.

To address this issue, researchers have explored semantic relationships between keywords with user-generated contents such as Wikipedia, DBpedia and manually-defined Web directories as valuable sources to generate non-obvious keywords (Amiri et al., 2008; Chen et al., 2008; Mirizzi et al., 2010). Wikipedia contains a large amount of clean information and conceptual knowledge on a wide spectrum of topics that can be utilized to suggest excessive long-tail keywords. Amiri et al. (2008) considered each query as an initial concept and tried to find related concepts from the Wikipedia collection. For each query keyword $q$, they grouped the retrieved documents by expectation maximization (EM) clustering algorithm and constructed a representative vector (i.e., keywords vector) for each cluster on the basis of the TFIDF scheme. The representative vector contains related keywords/concepts and the relationship weights between these keywords and $q$. Based on these vectors, a contextual graph was created to suggest a set of keywords/concepts that are more similar to the query keywords. DBpedia dataset is a community effort on the basis of Wikipedia to extract and store structured information in an RDF dataset that supports sophisticated queries. Mirizzi et al. (2010) presented a system to generate semantic tags through exploiting semantic relations from the DBpedia dataset. The similarity between each pair of resources in the DBpedia graph is calculated by querying external information sources (e.g., search engines and social bookmarking systems) and exploiting textual and link analysis in DBpedia. They used a hybrid ranking algorithm to rank keywords and expand queries formulated by users, and proved the validity of their algorithm by comparing with other RDF similarity measures (i.e., Algo2, Algo3, Algo4 and Algo5). By exploiting the semantic knowledge in the concept hierarchy built based on a manually-defined Web directory, Chen et al. (2008) proposed a keyword generation method. In more detail, it first matched a given seed keyword with one or several relevant concepts, and then these concepts were used together with the concept hierarchy to enrich the meaning of the seed keyword, finally a set of keywords was suggested by taking advantage of the conceptual information based on the similarity function of a variant of the Jaccard coefficient. However, similarity-based methods with semantic relationships suffer from the low concept coverage, and potentially lead to a decrease in the conversion rate, because low-cost keywords may be too far away from the initial seeds and thus fail to attract the target users.

Other sources such as co-bidding information and ads database have also been exploited in similarity-based keyword generation (Chang et al., 2009; Sarmento et al., 2009). Assuming that a set of bid keywords under the same ad is associated with a similar hidden intent, Chang et al. (2009) constructed a point-wise mutual information feature vector for each keyword $PMI(k) = (pmi_{k_1}, pmi_{k_2}, ..., pmi_{k_n})$, where $pmi_{k_i}$ is the pointwise mutual information between keyword $k$ and co-bidded keyword $k_i$. Then given an ad consisting of $n$ keywords, they ranked the suggested keywords in the ad network through summing the cosine similarity between the PMI feature vectors, i.e., $sim(k, k_i) = cosine(PMI(k), PMI(k_i))$ for $i = [1, ..., n]$. Similarly, assuming that advertisers associate inter-changeable keywords into the same ad, Sarmento et al. (2009) constructed a keyword synonymy graph by computing pairwise similarity between all co-occurrence vectors, and suggested relevant and non-obvious keywords by ranking candidate keywords through a function considering both the overlap and the average similarity. Through performing online comparisons of the proposed method with another keyword generation method used in the largest Portuguese Web advertising broker, they showed that advertisements containing keywords generated by the proposed method often have a superior performance in terms of click-through rate, which in turn resulted in a potential revenue increase.

### 3.2.3 Multivariate Models

Keyword generation methods based on similarity measures are heavily confined to the local space defined by seed keywords in that they rely on statistical or semantic relationships between keywords. Moreover, in many cases, keywords generated by similarity-based methods might fail to capture search users' real intents that are the most important to advertisers. To address these issues, multivariate models have been used to use rich latent information to facilitate keyword generation. Multivariate models can help advertisers generate a long tail of candidate keywords that are relevant and occupy a large fraction of the total traffic, by exploring potential factors and hidden topics affecting the performance of candidate keywords. Notations used in multivariate models for keyword generation are presented in Table 3c.

Table 3c. Notations used in multivariate models for keyword generation

| Terms | Definition |
|---|---|
| $y$ | A binary variable indicating whether a candidate keyword is relevant to a given seed keyword |
| $x$ | A vector of input features ($\bar{x}$) associated with a candidate keyword |
| $\bar{w}$ | A weight learned for an input feature in $x$ |
| $situation$ | Situation features specific to a targeted scene |
| $count(k, situation)$ | The number of times that $k$ occurs in $situation$ |
| $count(k)$ | The collection frequency of $k$ |
| $count(situation)$ | The number of keywords in scenes mapped to $situation$ |
| $U$ | A set of items (e.g., URLs) |
| $R_{|K|\times|U|}$ | A keyword-item rating matrix |
| $R_u$ | The $u$-th column of $R_{|K|\times|U|}$ |
| $R_{k,u}$ | The rate of $k$ on item $u$ |
| $Q$ | A binary column vector with $|K|$-dimensions where it is equal to 1 if the keyword on the $k$-th position is the seed keyword |
| $sim$ | A similarity vector |
| $\pi_m$ | A vector of indexed Web pages in the $m$-th iteration |
| $\pi_i^{(m)}$ | A score of the $i$-th Web page |
| $G$ | A row-normalized adjacency matrix of the link graph, and $G_{j,i} = \frac{1}{o_i}$ |

| $o_i$ | The out-degree of $i$-th Web page |
|---|---|
| $\mathbf{\Lambda}$ | A damping vector biased to a certain topic |
| $\mathbf{C}$ | A content damping vector |
| $\vartheta_1, \vartheta_2$ | Parameters controlling the impact of the content relevant score and the advertisement relevant score |
| $g$ | A row-normalized Wikipedia graph link $n \times n$ matrix |
| $topic$ | A topic |
| $n_{topic}$ | The number of topics |
| $n_d$ | The number of documents |
| $\varphi^{(topic)}$ | The topic distribution for document |
| $\varphi^{(k)}$ | The keyword distribution for topic |
| $\beta^{(topic)}, (\beta^{(k)})$ | Hyper parameters of Dirichlet distributions |
| $k_0$ | A seed keyword |
| $k_0.Cand$ | The set of candidate keywords for $k_0$ |
| $I$ | A set of keywords consisting of $k_0$ and $k_0.Cand$ |
| $k_i.AK$ | A list of associative keywords of $k_i$ |
| $k_i.Profile$ | A corresponding characteristic profile of $k_i$ |
| $k.vol$ | The search volume of $k$ |
| $Corpus$ | All characteristic profiles of keywords in $I$ |
| $\eta_j$ | The total times that the title keyword doesn't appear in the $j$-th component of a Wikipedia article |
| $\alpha_j$ | The importance of a component where a given keyword appears |
| $\theta_j$ | A random variable obeying an Beta distribution, denoting the unimportance of a component, i.e., $\theta_j = 1 - \alpha_j$ |
| $\alpha'_j, \beta'_j$ | The shape parameters for the Beta distribution of $\theta_j$ |
| $KWW(k)$ | The weight for a given keyword |
| $TFIDF_s(k)$ | The importance of a keyword presented in the abstract |
| $TFIDF_c(k)$ | The importance of a keyword presented in the content |
| $TFIDF_d(k)$ | The importance of a keyword presented in the main text |
| $TFIDF_i(k)$ | The importance of a keyword presented in the information box |
| $|AT(k)|$ | A variable indicating whether a keyword is in the anchor text |

In the literature on keyword generation, five multivariate models, namely logistic regression, collaborative filtering, topic-sensitive PageRank, latent Dirichlet allocation and hierarchical Bayesian have been employed to explore effects of various keyword characteristics in the generation process.

**(1) Logistic regression (LR).** LR is a basic learning technique which treats keyword generation as a binary classification problem. LR learns a vector of weights for input features and returns the estimated probability of whether a candidate phrase is relevant (Yih et al., 2006; Wu & Bolivar, 2008; Berlt et al., 2011; Lee et al., 2009). Typically, LR is used in keyword generation in the following form:

$$P(y = 1 | x = \bar{x}) = \frac{\exp(\bar{x} \cdot \bar{w})}{1 + \exp(\bar{x} \cdot \bar{w})}, \quad (4)$$

where $y$ is a binary variable, whose value is equal to 1 if the candidate keyword is relevant, otherwise 0; $x$ is a vector of input features ($\bar{x}$) associated with a candidate keyword; and $\bar{w}$ is a weight that the LR model learns for an input feature in $x$. The generated keywords are ranked by the estimated probability.

LR was used in finding advertising keywords from Web pages by Yih et al. (2006). Specifically, it takes $y$ as a binary variable under the monolithic selector; while under the decomposed selector, a phrase is decomposed into individual words and each word is tagged with one of the five labels (i.e., beginning, inside, last, unique and outside), and then five estimated possibilities are returned, i.e., $P(y = i|\bar{X} = \bar{x}) = \exp(\bar{x} \cdot \bar{w}_i) / \sum_{j=1}^{5} \exp(\bar{x} \cdot \bar{w}_j)$. The probability of a phrase is calculated by multiplying the five probabilities of its individual words being the correct label of the sequence. Experiments illustrated that LR significantly outperforms baseline methods (e.g., the TFIDF model, an extended TFIDF model with learned weights and a domain-specific keyword extraction method).

Subsequent studies used LR for different advertising scenarios. Following Yih et al. (2006), Wu & Bolivar (2008) added HTML features and proprietary data features (e.g., leaf and root category entropy) for the particular website (i.e., eBay) in the LR model, and constructed a candidate category vector for each keyword to resolve the keyword ambiguity problem. In order to find relevant keywords from online video contents, Lee et al. (2009) took into account not only within-document term features (e.g., TF-IDF scores), but also situation features specific to a targeted scene to train a LR model, i.e., pointwise mutual information (PMI) based on the co-occurrence information between a term $k$ and situation $situation$, $PMI(k, situation) \approx count(k, situation)/count(k)count(situation)$, where $count(k, situation)$ is the number of times that $k$ occurs in situation $situation$, $count(k)$ and $count(situation)$ are the collection frequency of $k$ and the number of words in scenes mapped to situation $situation$, respectively. Experiments showed that the scene-specific features are potentially useful to improve the performance of keyword extraction in video advertising. Instead of directly asking humans to evaluate the relevance of candidate keywords, Berlt et al. (2011) reduced the training cost of the LR-based keyword generation model by taking experts' evaluation on the relevance of advertisements for the page where the keyword candidate is extracted from. Experiments showed that their ad-collection-aware approach could yield significant gains without dropping precision values. However, the LR-based methods can only generate keywords from particular pages, while missing information from the similar pages. Besides websites and Web pages, advertisement databases and search click logs have been used by Bartz et al. (2006) to examine the performance of logistic regression, and results showed that biddedness data in advertisement databases can provide better precision.

**(2) Collaborative filtering (CF)**. CF is a classic recommendation algorithm. Suppose that there are a keyword set $K$ and an item (e.g., URL) set $U$. Following Bartz et al. (2006), the keyword-item rating matrix $\boldsymbol{R}_{|K|\times|U|}$ in keyword generation is given as

$$\boldsymbol{R}_{|K|\times|U|} = \begin{bmatrix} R_{1,1} & R_{1,2} & \ldots & R_{1,u} & \ldots & R_{1,|U|} \\ R_{2,1} & R_{2,2} & \ldots & R_{2,u} & \ldots & R_{2,|U|} \\ \vdots & \vdots & \vdots & \vdots & & \vdots \\ R_{k,1} & R_{k,2} & \cdots & R_{k,u} & \cdots & R_{k,|U|} \\ \vdots & \vdots & \vdots & \vdots & & \vdots \\ R_{|K|,1} & R_{|K|,2} & \cdots & R_{|K|,u} & \cdots & R_{|K|,|U|} \end{bmatrix}, \quad (5)$$

where $R_{k,u}$ denotes the rate of keyword $k$ on item $u$. Each keyword can be represented as a $|U|$-dimensional vector and the similarity between two rating vectors can be measured with the cosine similarity

$$cos(k, k') = \frac{R_k R_{k'}}{\|R_k\|_2 \|R_{k'}\|_2} = \frac{\sum_u R_{k,u} R_{k',u}}{\sqrt{\sum_u R_{k,u}^2}\sqrt{\sum_u R_{k',u}^2}}. \quad (6)$$

Then the keyword-item rating matrix could be converted to a keyword-term similarity matrix based on the cosine similarity, and the Top-k most similar keywords are ranked for keyword generation. Specifically, keywords and URLs in the search logs are extracted to construct a term-URL rating matrix $\boldsymbol{R}_{|K|\times|U|}$ whose rate is the number of times that a user searched for that keyword and clicked on that URL; then a column vector $\boldsymbol{Q}$ with $|K|$-dimensions was created, whose value is 0-1 binary, where it is equal to 1 if the keyword in the $k$-th position is the seed keyword. The similarity vector was calculated as $\boldsymbol{sim} = \sum_u 1(\boldsymbol{R}_u) \cos(\boldsymbol{R}_u, Q)$, where $\boldsymbol{R}_u$ is the $u$-th column of matrix $\boldsymbol{R}_{|K|\times|U|}$, and $1(\boldsymbol{R}_u)$ is an 0-1 binary indicator vector with $|K|$-length and contains 1 for every non-zero entry of $\boldsymbol{R}_u$. Finally, the keywords were ranked in descending order of indexes in $\boldsymbol{sim}$.

Based on advertisement databases and search click logs, Bartz et al. (2006) examined the performance of a CF model with respect to generating relevant keywords, starting from a set of seed keywords describing an advertiser's products or services, and found that the standard collaborative filtering framework has statistically equal performance with the logistic regression with a set of selected features.

**(3) Topic-sensitive PageRank (TSPR)**. Topic-sensitive PageRank extends PageRank by

allowing the iteration process to be biased to a specific topic. The main idea of the PageRank algorithm is to propagate the quality score of a Web page to its out-links and obtain the static quality of Web pages by performing a random walk on the link graph. Following Haveliwala (2003), the topic-sensitive PageRank algorithm is defined as

$$\boldsymbol{\pi}_{m+1} = (1 - \vartheta)\boldsymbol{\Lambda} + \vartheta G \boldsymbol{\pi}_m, \qquad (7)$$

where $\boldsymbol{\pi}_m = [\pi_1^{(m)}, \ldots, \pi_n^{(m)}]^T$ is a vector of the indexed Web pages in the $m$-th iteration, $\pi_i^{(m)}$ is the score of the $i$-th Web page; the matrix $G$ is a row-normalized adjacency matrix of the link graph, and $G_{j,i} = \frac{1}{o_i}$, $o_i$ is the out-degree of the $i$-th Web page; $\boldsymbol{\Lambda}$ is the damping vector biased to a certain topic, where $i$-th element in $\boldsymbol{\Lambda}$ is the relevance of the $i$-th Web page to the topic.

The propagation process of the topic-sensitive PageRank algorithm is biased by the damping vector $\boldsymbol{\Lambda}$ in each iteration. After iterations, Web pages with higher damping values propagate higher scores to their neighbors in the link graph.

In order to simultaneously generate keywords valuable for advertising from short-text Web pages, Zhang et al. (2012a) combined the content bias and advertisement bias into the propagation process of topic-sensitive PageRank, given as

$$\boldsymbol{R}_{m+1} = \vartheta_1 \boldsymbol{C} + \vartheta_2 \boldsymbol{A} + (1 - \vartheta_1 - \vartheta_2) g \boldsymbol{R}_m, \qquad (8)$$

where $\boldsymbol{C}$ is the content damping vector (i.e., the vector of the relevance between a set of Wikipedia entities and the target Web page) obtained through a regression based on Yih et al. (2006), and $\boldsymbol{A}$ is the advertisement damping vector obtained through calculating the frequency of each entity in the text of an advertisement; $\vartheta_1$ and $\vartheta_2$ are the parameters controlling the impact of the content relevant score to the content-sensitive PageRank value and the advertisement relevant score to the advertisement-sensitive PageRank value, respectively; $g$ is the row-normalized Wikipedia graph link $n \times n$ matrix, where $n$ stands for the size of the entity set. The TSPR-based keyword generation method can generate highly relevant keywords that don't occur on the target Web page, and can yield a significant improvement in precision over baseline methods (i.e., TF counting, supervised learning).

**(4) Latent Dirichlet allocation (LDA)**. Users' search intentions can be captured timely through exploring users' query topics hidden in query logs, which are valuable for commercial advertising and helpful in keyword generation (Qiao et al., 2017). Latent Dirichlet allocation (LDA) is a generative probabilistic model for a collection of documents, assuming that each document is represented as random mixtures over latent topics and each topic is characterized by a distribution over words (Blei et al., 2003).

Given a corpus consisting of $n_{topic}$ topics over $n_d$ documents, and each document contains $n_{d,k}$ keywords, let $\varphi^{(topic)}$ and $\varphi^{(k)}$ denote the topic ($topic$) distribution for document and the keyword ($k$) distribution for topic, respectively. Both $\varphi^{(topic)}$ and $\varphi^{(k)}$ obey Dirichlet distributions with hyper parameters $\beta^{(topic)}$ and $\beta^{(k)}$. Given the seed keyword $k_0$, Qiao et al. (2017) generated a set of candidate keywords ($k_0.Cand$) by analyzing indirect associations between $k_0$ and keywords in query logs. Let $I$ denote a keyword set consisting of seed keyword $k_0$ and candidate keywords $k_0.Cand$, i.e., $I = \{k_0\} \cup k_0.Cand$. Each keyword $k_i \in I$ has a list of associative keywords ($k_i.AK$), and the list of associative keywords constitutes a characteristic profile for $k_i$, i.e., $k_i.Profile = \{(k, k.vol)|k \in k_i.AK\}$, where $k$ is a keyword in $k_i.AK$ and $k.vol$ is its search volume. Taking each characteristic profile ($k_i.Profile$) as a document, all characteristic profiles of keywords in $I$ can be collected as a corpus, which is denoted as $Corpus = \{k_i.Profile|k_i \in I\}$.

In keyword generation, LDA interprets the characteristic profile of $k_i$ (i.e., $k_i.Profile$) as a multinomial distribution $Mult(\varphi^{(topic)})$ over a set of topics and each topic is assigned a multinomial distribution $Mult(\varphi^{(k)})$ over keywords in $Corpus$. Then the generation probability of a keyword $k$ in the $Corpus$ can be obtained through a process that samples a topic ($topic$) from $Mult(\varphi^{(topic)})$ specific with a characteristic profile ($k_i.Profile$) and subsequently samples a keyword ($k$) from $Mult(\varphi^{(k)})$ associated with $topic$, which is formulated as

$$P(k|k_i.Profile, \beta^{(topic)}, \beta^{(k)})$$
$$= \sum_{topic} P(k|topic, \beta^{(k)}) P(topic|k_i.Profile, \beta^{(topic)})$$
$$= \iint \sum_{topic} P(k|\varphi^{(k)}) P(topic|\varphi^{(topic)}) Dir(\varphi^{(topic)}|k_i.Profile, \beta^{(topic)}) d\varphi^{(topic)}$$
$$Dir(\varphi^{(k)}|topic, \beta^{(k)}) d\varphi^{(k)}. \qquad (9)$$

The sampling process discussed above is repeated $k.vol$ rounds for each keyword in a characteristic profile. Then the generation probability of the observed $Corpus$ can be obtained by repeatedly applying the sampling process to all the characteristic profiles, which is given as

$$P(Corpus|\beta^{(topic)}, \beta^{(k)}) = \prod_{k_i \in I} \prod_{k \in k_i.profile} \prod_{1 \leq r \leq k.vol} P(k|k_i.profile, \beta^{(topic)}, \beta^{(k)}). \quad (10)$$

Then Gibbs sampling was used to estimate parameters $\varphi^{(topic)}$ and $\varphi^{(k)}$ as well as the latent variable $topic$ by maximizing $P(Corpus|\beta^{(topic)}, \beta^{(k)})$. After the parameter estimation, each keyword can be projected into a topic distribution.

Considering the fact that two keywords related to similar topics might be competitive with each other due to market overlaps, Qiao et al. (2017) utilized topic distributions of keywords in query logs to infer competitive relationships between keywords. The mined topic structure is further combined into a factor graph model to extract a set of competitive keywords. The LDA-based method can generate competitive keywords which rarely co-occur with the seed keyword in queries.

On video sites with user-generated clips (e.g., YouTube), the text data is typically short. A video comprises a small number of hidden topics, which can be represented as keyword probabilities. Thereby, a video's text can be generated from some distribution over those topics. Welch et al. (2010) explored an LDA-based keyword generation method for video advertising by mining a range of short-text sources associated with videos. Compared to statistical n-gram keyword generation methods, the LDA-based method performed better when a limited amount of text data is available, and can substantially improve the matching relevance and the profitability of generated keywords.

**(5) Hierarchical Bayesian (HB)**. HB is a statistical model formed in multiple hierarchical structures which can be used to estimate the harmonic parameters in the keyword weighting formula for keyword generation. Given that a set of articles has been randomly selected from the corpus (e.g., Wikipedia articles), HB calculates the weight based on the importance of a keyword from the components of the Wikipedia article. Let $\eta_j$ be the total times that the title keyword doesn't appear in the $j$-th component, and $\theta_j$ be a random variable obeying a Beta distribution, i.e., $\theta_j \sim B(\alpha'_j, \beta'_j)$, based on the sample data $x_j$. Following Nie et al. (2019), the posterior joint probability distribution $p(\theta_j, \alpha'_j, \beta'_j | \eta_j)$ is

$$p(\theta_j, \alpha'_j, \beta'_j | \eta_j) = \frac{p(\eta_j|\theta_j,\alpha'_j,\beta'_j)p(\theta_j|\alpha'_j,\beta'_j)p(\alpha'_j,\beta'_j)}{p(\eta_j)} \propto p(\eta_j|\theta_j,\alpha'_j,\beta'_j)p(\theta_j|\alpha'_j,\beta'_j)p(\alpha'_j)p(\beta'_j). \quad (11)$$

By using HB, the weight of each keyword extracted from articles in the corpus can be calculated and used to decide its priority in the candidate set. Nie et al. (2019) presented a keyword generation method taking advantage of the rich link structure of Wikipedia's entry articles. Starting with a few seed keywords, the proposed method generates keywords in an iterative way, until a threshold is reached which balances the tradeoff between coverage and relevance of the generated keyword set. In the

keyword generation process, the weight of a keyword in an article was calculated as $KWW(k) = \alpha_1 TFIDF_s(k) + \alpha_2 TFIDF_c(k) + \alpha_3 TFIDF_d(k) + \alpha_4 TFIDF_i(k) + \alpha_5 |AT(k)|$, where $KWW(k)$ is the weight for a given keyword; $TFIDF_s(k)$, $TFIDF_c(k)$, $TFIDF_d(k)$ and $TFIDF_i(k)$ measure the importance of a keyword occurring in abstract, content, main text, information box, respectively; $|AT(k)|$ indicates whether a keyword is in the anchor text; and $\alpha_j$ ($\alpha_j = 1 - \theta_j, j = 1, \cdots, 5$) is the importance of a component where the keyword appears, which can be estimated by using the hierarchical Bayesian keyword weighting method. The HB-based method performs at a superior level with respect to both coverage and relevance.

### 3.2.4 Linguistics processing

Multivariate methods are incapable of generating semantically relevant keywords that don't contain or co-occur with the seed keywords. Moreover, multivariate methods suffer from the problem of topic drift which might generate keywords with little clicks and conversions and thus produce a massive overhead. Thus, more syntactic and semantic analysis are needed to capture users' real search intentions in keyword generation. Keyword generation based on linguistics processing helps find more semantically related and profitable keywords. Notations used in linguistics processing models for keyword generation are presented in Table 3d.

Table 3d. Notations used in linguistics processing models for keyword generation

| Terms | Definition |
| --- | --- |
| $k$ | $\boldsymbol{k} = \{k_{(1)}, k_{(2)}, \ldots, k_{(n)}\}$, denoting the bags of words of keyword $k$ |
| $landing$ | A landing page |
| $\boldsymbol{landing}$ | $\boldsymbol{landing} = \{landing_{(1)}, landing_{(2)}, \ldots, landing_{(m)}\}$, denoting the bags of words of $landing$ |
| $translation(landing_{(j)}|k_{(i)})$ | A translation table denoting the likelihood of $landing_{(j)}$ being generated from $k_{(i)}$ |
| $w_j$ | A weighting parameter for $landing_{(j)}$ |
| $mapping$ | A mapping from word(s) to sense(s) |
| $Senses_{DIC}(k_i)$ | The set of senses encoded in a dictionary $DIC$ for $k_i$ |
| $q$ | A query keyword |
| $\boldsymbol{q}$ | $\boldsymbol{q} = \{q_{(1)}, q_{(2)}, \ldots, q_{(n)}\}$, denoting a set of query keywords after sampling $n$ times |

In the following, we go through four linguistics processing methods in keyword generation, namely, translation model, word sense disambiguation, relevance-based language models and heuristics-based methods.

**(1) Translation model (TM)**. Translation model is originally presented in statistical machine translation literature to translate text from one type of natural languages to another (Brown et al., 1993). In keyword generation, translation model learns translation probabilities from keywords to landing pages through a parallel corpus, and bridges the vocabulary mismatch by giving credits to words in a phrase that are relevant to the landing page but do not appear as part of it. By treating keywords and landing page as bags of words, i.e., $\boldsymbol{k} = \{k_{(1)}, k_{(2)}, ..., k_{(n)}\}$ and $\boldsymbol{landing} = \{landing_{(1)}, landing_{(2)}, ..., landing_{(m)}\}$, for each $k_{(i)} \in \boldsymbol{k}$, the probability of the translation probability from the keyword to the landing page $P(\boldsymbol{landing}|\boldsymbol{k})$ can be estimated as

$$P(\boldsymbol{landing}|\boldsymbol{k}) \propto \prod_j \sum_i translation(landing_{(j)}|k_{(i)}), \qquad (12)$$

where $translation(landing_{(j)}|k_{(i)})$ is the translation table, i.e., the probability that characterizes the likelihood of a word in a landing page being generated from a word in a keyword.

Ravi et al. (2010) proposed a keyword generation method by combining a translation model with language models to produce highly relevant well-formed phrases. Specifically, the probability $P(\boldsymbol{landing}|\boldsymbol{k})$ was modified by associating a weight $w_j$ for all $landing_{(j)} \in \boldsymbol{landing}$ with respect to different features (e.g., a higher weight for words with HTML tags), i.e., $P(\boldsymbol{landing}|\boldsymbol{k}) \propto \prod_j \left( \sum_i translation\ (landing_{(j)}|k_{(i)}) \right)^{w_j}$, then the keyword language model $P(\boldsymbol{k})$ was instantiated with a bigram model capturing most of useful co-occurrence information and smoothed by a unigram model. Based on the Bayes' law, the probability $P(\boldsymbol{k}|\boldsymbol{landing})$, i.e., the likelihood of a keyword given the landing page, can be estimated to rank the keywords, i.e., $P(\boldsymbol{k}|\boldsymbol{landing}) \propto P(\boldsymbol{landing}|\boldsymbol{k}) P(\boldsymbol{k})$. Experiments based on a realworld corpus of landing pages and associated keywords showed that the TM-based method outperformed significantly over a method based purely on text extraction, and could generate many human-crafted keywords.

**(2) Word sense disambiguation (WSD)**. WSD is the ability in computational linguistics to identify which sense (meaning) of a word is used in a particular context (Navigli, 2009). Through viewing a text as a sequence of words $(k_1, k_2, ..., k_n)$, WSD can be defined as a task of assigning the appropriate sense(s) to word(s) in the text, i.e., to identify a mapping ($mapping$) from word(s) to sense(s), such that $mapping(i) \subseteq Senses_{DIC}(k_i)$, where $Senses_{DIC}(k_i)$ is the set of senses encoded in a dictionary $DIC$ for word $k_i$, and $mapping(i)$ is that subset of the senses of $k_i$ which are appropriate in the text. In online advertising and SSA, in order to exclude the display of an advertisement from a population of non-target audiences, it is necessary to identify negative keywords. For example, Scaiano & Inkpen (2011) proposed a method to automatically identify negative keywords, using Wikipedia as a sense inventory and an annotated corpus. Specifically, they searched all links containing the seed keywords and collected all destination pages to find all senses for each keyword, then generated context vectors for each sense by tokenizing each paragraph containing a link to the sense being considered. In this process, all words were recorded and counted as a dimension in the context vector. After identifying the intended sense and creating a broad-scope intended-sense list, negative keywords are identified through finding words from the context vectors of the unintended senses with high TFIDF. The WSD-based method could find keywords strongly correlated with negative topics and improve the performance of advertising campaigns.

**(3) The relevance-based language (RBL) model**. The RBL model is used to determine the probability of observing a keyword $k$ in a collection of documents, i.e., $P(k|DOC)$ (Lavrenko & Croft, 2017). Given a query keyword $q$ and a large collection of documents ($DOC$), both $q$ and each document are represented as a sequence of words. Each document in $DOC$ is related to $q$ and there is no training data about which document in $DOC$ is related to $k$. The relevance model is referred to the underlying mechanism that determines the probability $P(k|DOC)$. We assume that both $q$ and documents related to $k$ can be sampled from $DOC$, but possibly by different sampling processes. After sampling $n$ times, we can observe query keywords $\boldsymbol{q} = \{q_{(1)}, q_{(2)}, \dots, q_{(n)}\}$. Thus, given such observations, we can estimate the conditional probability of observing $k$ as

$$P(k|DOC) \approx P(k|q_{(1)}q_{(2)}\dots q_{(n)}) = \frac{P(k, q_{(1)}q_{(2)}\dots q_{(n)})}{P(q_{(1)}q_{(2)}\dots q_{(n)})}. \quad (13)$$

Jadidinejad & Mahmoudi (2014) proposed a keyword generation method based on a modified relevance-based language model, with Wikipedia as the knowledge base. First, by capitalizing Wikipedia's disambiguation pages, different semantic groups for a given ambiguous query were extracted. Second, appropriate semantic groups were selected based on user's intent, and an initial list of candidate keywords were generated by tracking bidirectional anchor links. Third, given a seed query and a collection of documents corresponding to candidate keywords from Wikipedia, the relevance-based language model was applied to estimate the probability of observing a keyword in documents, which helped measure the relevance between candidate keywords to the seed query. The RBL-based method is language independent, well-grounded with expert keywords and computationally efficient.

**(4) Heuristics-based method**. Consumers with a high conversion probability tend to use an online store's internal search engine (Ortiz-Cordova et al., 2015). Scholz et al. (2019) proposed a heuristics-based method to extract keywords from an online store's internal search logs. Specifically, a set of candidate keywords identified from the internal search and other sources (e.g., Thesauri) was enriched with other keyword generation tools and filtered according to the monthly search volume in Google, then keywords whose internal search volumes are higher than a predefined threshold were included into the target set. This heuristics-based method can substantially increase the number of profitable keywords and the conversion rate, and in the meanwhile, decrease the average cost per click.

### 3.2.5 Machine learning

The performance of linguistics processing is influenced by the quality of sources (e.g., Web pages, texts, dictionaries), which might make suggested keywords deviate from search users' real intentions. Machine learning can automatically learn and produce more accurate and reliable results by consuming a large amounts of data accumulated in online advertising, executing feature engineering without little interference of humans and capturing more rich behavioral information and structural relationships with high-order representation. Thus, machine learning models can generate large sets of relevant keywords by capturing the intention of users and processing the up-to-date information sources. Notations used in machine learning models for keyword generation are presented in Table 3e.

Table 3e. Notations used in machine learning models for keyword generation

| Terms | Definition |
|---|---|
| $\xi_n$ | A random variable describing the position of random walk after $n$ steps |
| $w_n$ | The $n$-th step of a random walk |
| $\tilde{c}$ | A concept that an advertiser is interested in |
| $l_k$ | The random variable related to a concept for query keyword $k$ |
| $l_u$ | The random variable related to a concept for URL $u$ |
| $\mathcal{T}$ | The probability of transiting to the absorbing null class node |
| $D$ | Dataset |
| $\hat{n}$ | A node in decision tree |
| $\hat{n}'$ | A child node of $\hat{n}$ |
| $C(\hat{n}|\Upsilon)$ | A set of child nodes of $\hat{n}$ |
| $D_{\hat{n}}$ | A subset of dataset $D$ at node $\hat{n}$ |
| $\Upsilon$ | The branching criterion |
| $G(\hat{n},\Upsilon)$ | The quality of the partition of $D_{\hat{n}}$ induced by $\Upsilon$ |
| $x_\Upsilon$ | An element of $x$ |
| $\mathcal{X}_\Upsilon$ | A set of unique values of categorical variable $x_\Upsilon$ |
| $\mathcal{X}_{\Upsilon,k}$ | The $k$-th value of $\mathcal{X}_\Upsilon$ |
| $\|\mathcal{X}_Y\|$ | The number of values in $\mathcal{X}_Y$ |
| $H$ | The entropy |
| $seq, seq'$ | A sequence |
| $seq_i\ (seq'_i)$ | The $i$-th element in sequence $seq(seq')$ |
| $seq\_id$ | The id of sequence $seq$ |
| $seq\_pattern$ | A sequential pattern |
| $\boldsymbol{K}_T$ | The union of the keyword matrix $[\boldsymbol{k}_{T,1}, \ldots, \boldsymbol{k}_{T,n}]^T$ and the target dataset $\{\boldsymbol{k}_{T,i}\}$ |
| $\boldsymbol{K}_L$ | The union of the keyword matrix $[\boldsymbol{k}_{L,1}, \ldots, \boldsymbol{k}_{L,m}]^T$ and $\{\boldsymbol{k}_{L,i}\}$, i.e., a subset of $\boldsymbol{K}_T$ that is chosen to be labeled |
| $\epsilon_i$ | The measurement error |
| $f_T$ | $f_T = [f(k_{T,1}), \ldots, f(k_{T,n})]^T$, denoting the function values on all the available data $K_T$ |
| $\sigma^2$ | The variance of a distribution |
| $\sigma^2 \boldsymbol{C}_f$ | A covariance matrix |
| $s_{ik}$ | A binary parameter indicating whether the $i$-th ad is subscribed with the keyword $k$ |
| $C_i$ | A binary variable indicating whether the $i$-th ad belongs to category $j$ |
| $\mu_{jk}$ | The mean of the prior distribution for the keyword subscription probability |
| $\tilde{r}_{ij}$ | The probability that the $i$-th ad belongs to category $j$ |
| $cand_{k_{n-1}}$ | A candidate set of keywords with one-step relevance to $k_{n-1}$ |
| $R_1(k_i, k_{i+1})$ | The one-step relevance of keyword pair $(k_i, k_{i+1})$ |
| $R_n$ | $R_n(k_0, k\|k_1, k_2, \ldots, k_{n-1})$, denoting the n-step relevance of $k_0$ and $k$ via the intermedia keywords $k_1, k_2, \ldots, k_{n-1}$ |
| $\boldsymbol{H}$ | $\boldsymbol{H} = (\boldsymbol{h}_1, \boldsymbol{h}_2, \ldots, \boldsymbol{h}_n)$, denoting hidden representations |
| $\boldsymbol{e}(k_t)$ | The embedding of keyword $k_t$ |
| $\boldsymbol{ct}_{t-1}$ | A context vector |
| $\boldsymbol{s}_t$ | A state vector |
| $\widetilde{K}$ | $\widetilde{K} = (\tilde{k}_1, \tilde{k}_2, \ldots, \tilde{k}_m)$, denoting a generated target keyword sequence |
| $domain_k, domain_{\tilde{k}}$ | The corresponding domain categories of $K$ and $\widetilde{K}$ |

In the literature, researchers have investigated several statistical learning methods to facilitate keyword generation in SSA, including random walk, decision tree, sequential pattern mining, active learning, Bayesian online learning, multi-step semantic transfer analysis and sequence-to-sequence learning.

**(1) Random walk (RW)**. RW can be used to describe a keyword generation path including a succession of random steps in the query-click graph extracted from search logs. In essence, clicks represent a strong association between queries and URLs (Yang & Zhai, 2022). Hence, advertisers can exploit the proverbial "wisdom of the crowds" to reconstruct query-click logs as a weighted bipartite graph $(K, U, \boldsymbol{R})$. Specifically, query keywords in $K$ and URLs in $U$ constitute the partitions of the graph, and the number of times that users issued query $k$ to the search engine and clicked on URL $u$ (i.e., $R_{k,u}$) can be regarded as weight on the edge $(k, u) \in R$. Suppose that $w_1, w_2, ..., w_n$ is a sequence of independent and identically distributed random variables, a random walk is a random process which describes a path consisting of a succession of random steps on some mathematical space (Xia et al., 2019). It can be denoted as $\{\xi_n, n = 0,1,2, ...\}$, where $\xi_n$ is a random variable describing the position of random walk after $n$ steps and $\xi_n = \xi_{n-1} + w_n, n \geq 1$, where $w_n$ denotes the step of the random walk.

Fuxman et al. (2008) formulated the keyword generation problem within a framework of Markov Random Fields and developed an RW-based algorithm with absorbing states to traverse a query-click graph. Given a concept $\tilde{c}$ that an advertiser is interested in, a seed set of URLs relevant to $\tilde{c}$ was constructed manually, which can be regarded as the representation of $\tilde{c}$. Let $l_k$ and $l_u$ denote the random variable related to $\tilde{c}$ for query keyword $k$ and URL $u$, respectively. The probability that a random walk starting from query $k$ will be absorbed at concept $\tilde{c}$, i.e., $P(l_k = \tilde{c})$, can be computed as

$$P(l_k = \tilde{c}) = (1 - \mathcal{T}) \sum_{u:(k,u) \in R} \frac{R_{k,u}}{\sum_{u:(k,u) \in R} R_{k,u}} P(l_u = \tilde{c}), \qquad (14)$$

where $\mathcal{T}$ is the probability of transiting to the absorbing null class node. The null class node is a node in a query-click graph whose probability (i.e., $P(l_k = \tilde{c})$ or $P(l_u = \tilde{c})$) is below a threshold, and the set of null class nodes define the boundary of the query-click graph. Similarly and recursively, for all URLs in the seed set, $P(l_u = \tilde{c}) = 1$; for other URLs (i.e., unlabeled URL) in search logs, the probability of a random walk that starts from URL $u$ and ends up being absorbed in concept $\tilde{c}$, i.e., $P(l_u = \tilde{c})$, can be computed as

$$P(l_u = \tilde{c}) = (1 - \mathcal{T}) \sum_{k:(k,u) \in R} \frac{R_{k,u}}{\sum_{k:(k,u) \in R} R_{k,u}} P(l_k = \tilde{c}). \qquad (15)$$

The processes defined by $P(l_k = \tilde{c})$ and $P(l_u = \tilde{c})$ iterated alternately until the convergence. Discarding $P(l_k = \tilde{c})$ and $P(l_u = \tilde{c})$ whose probabilities lower than a predefined threshold, we can obtain $P(l_k = \tilde{c})$, the probability that query $k$ belongs to seed concept $\tilde{c}$, for every query in the set $K$. we can obtain $P(l_k = \tilde{c})$, i.e., the probability that $k$ will be absorbed at $\tilde{c}$, for every query keyword in search logs, which can be regarded as the relevance between $k$ and $\tilde{c}$. The RW-based can generate a large amount of high-quality keywords with minimal effort from advertisers.

**(2) Decision tree (DT)**. DT is a flowchart-like structure where paths from the root to leafs represent classification rules. Given a dataset $D = \{(x, y)\}$ with $x$ being a feature vector of keywords and $y$ being the label of relevance, DT represents a recursive partition of $D$ such that (a) each node of the tree stores a subset of $D$ with the root node storing $D$, (b) the subset $D_{\hat{n}}$ at node $\hat{n}$ is the union of the mutually disjoint subsets stored at its child nodes ($\hat{n}'$), i.e., $\{D_{\hat{n}'}|\ \hat{n}' \in C(\hat{n}|Y)\}$ forms a partition of $D_{\hat{n}}$ where $C(\hat{n}|Y)$ denotes the set of child nodes of $\hat{n}$, and (c) the partition is determined by a branching criterion $Y$. The optimal DT is built by recursively identifying the locally optimal branching criterion at each node starting from the root node while subjecting to some stopping as well as pruning criteria. Specifically, at node $\hat{n}$, the optimal branching criterion is

$$\theta_{\hat{n}}^* = argmax_Y\ G(\hat{n}, Y), \qquad (16)$$

where $G(\hat{n}, Y)$ measures the quality of the partition of $D_{\hat{n}}$ induced by $Y$.

In the decision tree-based scheme, GM et al. (2011) developed a keyword generation approach to learn the website-specific hierarchy from the (Web page, URL) pairs of a website, and keywords are populated on nodes of the induced hierarchy via successive top-down and bottom-up iterations. Human evaluations showed that their method outperformed previous approaches by Broder et al. (2007) and Anagnostopoulos et al. (2007) in terms of relevance. In keyword generation, as specified by GM et al. (2011), an instance $(x, y)$ corresponds to a Web page, where $x$ is a vector of features extracted from the URL of the Web page, and $y$ is the cluster of Web pages with similar contents. As an example, consider a Web page with the URL "*www.examplewear.com/exampleshop/product.php?view=detail& group=shoes&dept=men*", from which four features can be extracted, i.e., $x_1 =$"product.php" (the name of the php script), $x_2 =$"detail" (the value of argument "view"), $x_3 =$"shoes" (the value of argument "group") and $x_4 =$"men" (the value of argument "group"). Each element of $x$, i.e., $x_Y$, is a categorical variable having a set of unique values denoted by $\mathcal{X}_Y$. For example, $\mathcal{X}_4 = \{$"men", "women", "outlet", "accessories"$\}$, which is constructed by going through all URLs and collecting values of the "dept" argument. For convenience, let $\mathcal{X}_{Y,k}$ denote the $k$-th value of $\mathcal{X}_Y$ assuming an arbitrary order, and $|\mathcal{X}_Y|$ denote the number of values in $\mathcal{X}_Y$. The branching criterion is to select a feature $Y$, according to which $D_{\hat{n}}$ is partitioned. The quality of the resulting partition is measured with the gain ratio metric, specified as follows.

$$G(\hat{n}, Y) = \frac{H(y|D_{\hat{n}}) - \sum_{k=1}^{|\mathcal{X}_Y|} P(x_Y = \mathcal{X}_{Y,k}|D_{\hat{n}}) H\left(y|D_{(\hat{n}, \mathcal{X}_{Y,k})}\right)}{H(x_Y|D_{\hat{n}})}, \qquad (17)$$

where $P(x_Y = \mathcal{X}_{Y,k}|D_{\hat{n}})$ is the estimated probability of having value $\mathcal{X}_{Y,k}$ on feature dimension $i$ for a data instance in $D_{\hat{n}}$, $H(y|D_{\hat{n}})$ is the entropy of labels in $D_{\hat{n}}$ measuring how impure (diversified)

Web pages in $\mathcal{D}_{\hat{n}}$ are in terms of assigned clusters, $H\left(y|D_{(\hat{n}, \mathcal{X}_{Y,k})}\right)$ is the entropy of labels in $D_{(\hat{n}, \mathcal{X}_{Y,k})}$, and $H(x_Y|D_{\hat{n}})$ is the entropy of feature $Y$ in $D_{\hat{n}}$ measuring the complexity of the partition. The gain ratio metric measures how much impurity (content dissimilarity) reduction can be achieved through a data space partition, and favors partitions with high impurity reduction but low partition complexity, which is helpful in preventing overfitting.

**(3) Sequential pattern mining (SPM)**. SPM aims to find frequent patterns from a set of sequences (Mabroukeh & Ezeife, 2010), which is used to find keywords in online broadcasting contents. Given a set of items (e.g., terms) $K = \{k_1, k_{2,\ldots}, k_{n_K}\}$, two sequences $seq = <seq_1 seq_2 \ldots seq_{n_{seq}}>$ and $seq' = <seq'_1 seq'_2 \ldots seq'_{n_{seq'}}>$, where $seq_i$ ($seq'_i$) is a subset of items $K$, if there exist integers $1 \leq j_1 < 2 < j_2 < \cdots < n_\alpha \leq j_{n_\alpha}$ making $seq'_1 \subseteq seq_{j_1}, seq'_2 \subseteq seq_{j_2}, \ldots, seq'_{n_\alpha} \subseteq seq_{j_{n_\alpha}}$, $seq'$ is called a subsequence of $seq$, i.e., $seq' \subseteq seq$. Given a sequence dataset $D$, i.e., a set of tuples $\langle seq\_id, seq \rangle$, where $seq$ is a sequence and $seq\_id$ is the id of $seq$, the support of a subsequence $\alpha'$ is the number of tuples in the dataset containing $seq'$, given as

$$support_D(seq') = |\{\langle seq\_id, seq \rangle | (\langle seq\_id, seq \rangle \in D) \wedge (seq' \in seq)\}|. \quad (18)$$

Given a positive integer $min\_support$ as the support threshold, a sequence $seq'$ is called a sequential pattern if $support_D(seq') \geq min\_support$.

In keyword generation from online community contents, Li et al. (2007) used sequential mining to discover language patterns (i.e., a sequence of frequent words around an extracted keyword). The Web has become a communication platform, where users spend a large amount of time on broadcasting and interactions with others in online communities, e.g., blogging, posting, chatting, etc. Online contents are composed of specific keywords, phrases and wordings associated with frequently changed topics in communities. Keywords are extracted once a sentence is matched with a pattern from online broadcasting contents and scored with the sum of the confidence of matched patterns, i.e., $\sum confidence(seq\_pattern) = \sum support_D(seq)/support_D(seq\_pattern\backslash <topic>)$, where $seq\_pattern\backslash <topic>$ is the remaining part of sequential pattern $seq\_pattern$ after $<topic>$ is removed. The process of sequential pattern mining and keyword extraction iterates and eventually generates a large number of keywords. Experiments showed that the proposed approach can find meaningful language patterns and reduce the cost of manual data labeling, compared with traditional statistical approaches that considered each word individually.

**(4) Active learning (AL)**. AL is a special type of machine learning where a learning algorithm actively queries users (or some information sources) to label new data points with the desired outputs under situations where unlabeled data is abundant but manual labeling is expensive. Transductive Experimental Design is an active learning approach which can be used to select candidate keywords (for labeling and training) that are hard to predict and representative for unlabeled candidates. Let $K_T$ denote the union of the keyword matrix $[k_{T,1}, ..., k_{T,n}]^T$ and the target dataset $\{k_{T,i}\}$, and $K_L$ denote the union of the keyword matrix $[k_{L,1}, ..., k_{L,m}]^T$ and a subset of $K_T$ that is chosen to be labeled (i.e., $\{k_{L,i}\}$). Define $f(x) = w^T k_L$ as the output function learned from the measure $y_i = w^T k_{L,i} + \epsilon_i, i = 1, ..., m,$ where $w$ is the weight vector, $\epsilon_i \sim N(0, \sigma^2)$ is measurement error and $y_i$ (label) is the binary relevance score. Let $f_T = [f(k_{T,1}), ..., f(k_{T,n})]^T$ be the function values on all the available data $K_T$. Then the predictive error $f_T - \hat{f}_T$ has the covariance matrix $\sigma^2 C_f$ with

$$C_f = K_T (K_L^T K_L + \mu I)^{-1} K_T^T. \quad (19)$$

The total predictive variance on the complete data set $K_T$ is given as

$$\sum_{i=1}^n E[(F(k_{T,i}) - f(k_{T,i}))^2] = \sigma^2 Tr(C_f). \quad (20)$$

The objective is to find a subset $K_L$ which can minimize the total predictive variance.

Users' relevance feedback is another type of valuable information source for profitable keyword generation. Wu et al. (2009) proposed an efficient interactive model based on an active learning approach called transductive experimental design using relevance feedback for keyword generation in SSA. Each keyword was represented using a characteristic document consisting of top-hit search snippets for a seed keyword. In a seed's characteristic document, top-n weighted terms were recommended as candidate keywords. The AL-based method could significantly improve the relevance of generated keywords.

**(5) Bayesian online learning (BOL)**. Bayesian online learning replaces the true posterior distribution with a simple parametric distribution, and defines an online algorithm by a repetition of two steps (i.e., an update of the approximate posterior when a new sample arrives and an optimal projection into the parametric family) (Opper & Winther, 1999). BOL is helpful to improve the computational efficiency when estimating the unknown variables based on a large data. In SSA, advertisers use a set of keywords to describe an advertisement. Let $K$ be the set of subscribed keywords in the $i$-th ad: if the $i$-th ad is subscribed with keyword $k \in K$, then $s_{ik} = 1$, else $s_{ik} = 0$. Assuming that the keyword vector of an ad is sampled from one or several ad categories, such as automobiles and travel, let $p(s_{ik} = 1 | C_i = j)$ denote the probability that the $i$-th ad is subscribed by $k$ when it belongs to category $j$. Then under the scheme of Bayesian online learning, each data point of ads is processed at a time, and the posterior distributions of the probability $p(s_{ik} = 1 | C_i = j)$ obtained after processing a data point are passed as the prior distributions for processing the next data point. Keywords can be generated to an advertiser based on keyword subscriptions of other advertisers. The probability of an unobserved keyword $k' \notin K$ that is implicitly related to the $i$-th ad can be given as

$$p(s_{ik'} = 1 | \{s_{ik}\}_{k \in K}) = \sum_{j=1}^{n} \tilde{r}_{ij} \mu_{jk}, \quad (21)$$

where $\tilde{r}_{ij} = p(C_i = j | \mathbf{s}_i)$ is the probability that the $i$-th ad belongs to category $j$, and $\mu_{jk}$ is the mean of prior distribution for the keyword subscription probability $p(s_{ik} = 1 | C_i = j)$.

Schwaighofer et al. (2009) provided an efficient Bayesian online learning algorithm to group advertisements into categories and applied the BOL algorithm to generate keywords. Experiments based on two advertisement datasets showed that the BOL-based algorithm is suitable for large scales of data streams because of its low computational cost.

**(6) Multi-step semantic transfer analysis (MTSTA)**. The MTSTA-based keyword generation can yield keywords based on both their direct and indirect relevance to the seed keywords via semantic transfer. Given a seed keyword $k_0$, for keyword $k$, if there exist $n-1$ intermedia keywords $k_1, k_2, \ldots, k_{n-1}$ satisfying the conditions that $k$ is in the candidate set of $k_{n-1}$ (i.e., $cand_{k_{n-1}}$), $k_{n-1}$ is in the candidate set of $k_{n-2}$ (i.e., $cand_{k_{n-2}}$), …, and $k_1$ is in the candidate set of $k_0$ (i.e., $cand_{k_0}$), then the n-step relevance of $k_0$ and $k$ via the intermedia keywords $k_1, k_2, \ldots, k_{n-1}$, can be defined as

$$R_n(k_0, k | k_1, k_2, \ldots, k_{n-1}) = R_1(k_0, k_1) \prod_{i=1}^{n-2} R_1(k_i, k_{i+1}) R_1(k_{n-1}, k), \quad (22)$$

where $R_1(k_0, k_1), R_1(k_i, k_{i+1}), R_1(k_{n-1}, k)$ are the one-step relevance of keyword pairs.

The MTSTA-based keyword generation finds keywords with multi-step relevance that is no less than a certain threshold in the query logs (Zhang & Qiao, 2018; Zhang et al., 2021). In order to explore keywords with indirect relevance, Zhang and his colleagues explored a MTSTA-based keyword generation method by iteratively conducting co-occurrence analysis to form a hierarchal multi-step relevance tree, and developed a pruning strategy to reduce the computational consumption in generating the transfer paths.

**(7) Sequence-to-sequence learning (Seq2Seq).** The encoder-attention-decoder framework based on Seq2Seq learning is an end-to-end approach to sequence learning that makes minimal assumptions on the sequence structure (Sutskever et al., 2014). In keyword generation, the encoder represents an input keyword sequence $K = (k_1, k_2, ..., k_n)$ with hidden representations $\boldsymbol{H} = (\boldsymbol{h}_1, \boldsymbol{h}_2, ..., \boldsymbol{h}_n)$, i.e.,

$$\boldsymbol{h}_t = GRU(\boldsymbol{h}_{t-1}, \boldsymbol{e}(k_t)), \qquad (23)$$

where GRU is gated recurrent unit (Chung et al., 2014) and $\boldsymbol{e}(k_t)$ is the embedding of keyword $k_t$.

The decoder updates state $\boldsymbol{s}_t$ as follows:

$$\boldsymbol{s}_t = GRU(\boldsymbol{s}_{t-1}, [\boldsymbol{ct}_{t-1}; \boldsymbol{e}(\tilde{k}_{t-1})]), \qquad (24)$$

where $\boldsymbol{ct}_{t-1}$ is the context vector defined as a weighted sum of the encoder's hidden states, and $\boldsymbol{e}(\tilde{k}_{t-1})$ is the embedding of a previously decoded keyword.

After obtaining the state vector $\boldsymbol{s}_t$, the decoder samples from the generation distribution and generates a keyword $\tilde{k}_t$:

$$\tilde{k}_t \sim P(\tilde{k}_t | \tilde{k}_1, \tilde{k}_2, ..., \tilde{k}_{t-1}, \boldsymbol{ct}_t) = softmax(\boldsymbol{w} \cdot \boldsymbol{s}_t). \qquad (25)$$

$\widetilde{K} = (\tilde{k}_1, \tilde{k}_2, ..., \tilde{k}_m)$ forms a sequence of generated keywords.

Zhou et al. (2019) proposed a keyword generator based on Seq2Seq learning to generate domain-specific keywords through estimating the probability $P(\widetilde{K}, domain_{\tilde{k}} | K, domain_k) = P(domain_{\tilde{k}} | K, domain_k) \prod_{t=1}^{m} (\tilde{k}_t | \tilde{k}_1, \tilde{k}_2, ..., \tilde{k}_{t-1}, domain_{\tilde{k}}, K, domain_k)$, where $domain_k$ and $domain_{\tilde{k}}$ are the corresponding domain categories of $K$ and $\widetilde{K}$, respectively. In addition, a reinforcement learning algorithm was developed to strengthen the domain constraint in the generation process. The Seq2Seq-based method could generate diverse, relevant keywords within the domain constraint.

However, statistical learning methods have some limitations, such as requiring a set of labelled keywords and the low efficiency in online computation.

### 3.3 Features used for Keyword Generation

In the literature, keyword generation methods have been proposed on the basis of five major information sources to extract keywords and relationships among them, which are described as follows.

(1) Websites and Web pages: The Web has become a vital place for firms to post advertisements and other commercial information (Thomaidou and Vazirgiannis, 2011). In the meanwhile, the richness of information sources on the Web entitles advertisers to build a domain-specific keyword pool. In particular, websites and Web pages can be used as a corpus of the source text to extract relevant keywords of interest for their online advertising campaigns. In this branch of keyword generation methods, meta-tags of Web pages are used as an important information feature. The meta-tag crawler sends one or more seed keywords to search engines and extracts a set of meta-tag keywords from Web pages in the organic list. Several popular online advertising tools (e.g., WordStream and Wordtracker) employ meta-tag crawlers to obtain a pool of meta-tag keywords and then based on it suggest relevant keywords for advertisers.

(2) Search users' query logs: User's query logs with search engines timely reflect their intents (Da et al., 2011), which are significantly valuable for commercial communications and advertising. This stream of keyword generation primarily utilizes statistical information of co-occurrence relationships among keywords mining from historical query logs.

(3) Search results snippets: One or several seed keywords are sent to search engines and resulting search result snippets are used to generate relevant keywords.

(4) Advertisement databases and advertisers' bidding data: Search advertisement databases and advertising logs such as bidding data are taken as inputs to obtain relevant keywords.

(5) Domain semantics and concept hierarchy: Keyword generation methods relying on query logs mining generally ignore the semantic similarity between keywords, thus fail to suggest keywords that don't explicitly contain seed keywords or have less co-occurrence with but are semantically related to seed keywords. To this end, the fifth category of keyword generation primarily focuses on the expansion of the keyword scope by taking advantage of conceptual hierarchies built manually or extracted either from vocabulary dictionaries/corpus (e.g., thesaurus dictionary, Wikipedia) or constructed by domain experts.

In the following, we explore features used in prior research in the five streams. Tables 4a-4e summarize input/features used in keyword generation in five research streams.

In the literature, a variety of features are used to represent keywords, which have great contributions to the effectiveness of keyword generation solutions. In keyword generation from websites and Web pages, from Table 4a, it is apparent that information retrieval oriented features are most widely used in keyword extraction from websites and Web pages. As reported by Yih et al. (2006), information retrieval oriented features and query log features are helpful for keyword generation, while linguistic features don't seem to work. Consequently, Berlt et al. (2011) adopted features extracted from the ad collection and GM et al. (2011) took the hierarchic URL tokens as features, while omitting linguistic features. However, Li et al. (2007) found that features of language patterns can help keyword generation, and Zhou et al. (2007) inserted features such as title and keyword importance into meta keywords vector to improve keyword generation. In addition to features from Web pages as in Yih et al. (2006), Wu and Bolivar (2008) explored features from the view of retailers (e.g., eBay). In keyword generation for video advertising, Lee et al. (2009) advocated features reflecting the targeted scene situation.

Table 4a. Input/Features for Keyword Generation from Websites and Web Pages

| Refs. | Features | | | | | | | | | | | | | | | | | | |
|---|---|---|---|---|---|---|---|---|---|---|---|---|---|---|---|---|---|---|---|
| | LF | CA | HY | MF | TI | URL | IRF | RL | SDL | LCP | QLF | RE | NRC | H1 | CO | CID | TD | LP | SF |
| Yih et al. (2006) | √ | √ | √ | √ | √ | √ | √ | √ | √ | √ | √ | | | | | | | | |
| Wu & Bolivar (2008) | | | | √ | √ | | √ | √ | | √ | √ | √ | √ | √ | | | | | |
| Berlt et al. (2011) | | √ | √ | √ | √ | √ | √ | √ | √ | √ | √ | | | | | | | | |
| GM et al. (2011) | | | | | | √ | √ | | | | | | | | | | | | |
| Zhou et al. (2007) | | | | √ | √ | √ | | | | | | | | | √ | √ | √ | | |
| Li et al. (2007) | | | | | | | | | | | | | | | | | | √ | |
| Lee et al. (2009) | √ | | | | | √ | √ | | √ | √ | | | | | | | | √ | √ |

Note: LF=Linguistic Features; CA=Capitalization; HY=Hypertext; MF=Meta related Features (e.g., meta section, meta keywords, meta description); TI=Title; URL=Uniform Resource Locator; IRF=Information Retrieval Oriented Features (e.g., TF, IDF, TF history, log value of TF and DF); RL=Relative Location (e.g., wordRatio, sentenceRatio, wordDocRatio); SDL=Sentence and Document Length; LCP=Length of the Candidate Phrase; QLF=Query Log Features (e.g., whether the word appears in the query log files as the first/interior/last word of a query keyword, whether the word never appears in any query log); RE=Root Entropy; NRC=The Number of Root Categories; H1=the Highest Section Level; CO=Co-occurrence; CID=Class ID (i.e., the category of Web pages or advertisements such as sports); TD=Text Descriptions (i.e., a detail description to the product, company or related matter of Web pages or advertisements); LP=Language Pattern; SF=Situation Features.

In online advertising, clicks demonstrate a strong relationship between queries and URLs. This makes query logs valuable information for keyword generation (Bartz et al., 2006; Fuxman et al., 2008), as illustrated in Table 4b. Meanwhile, joint search demand and keyword search demand are informative and helpful in keyword expansion and competitive strategy development (Qiao et al., 2017). Moreover, semantic and domain-specific information entitles to generate keywords that aren't present in the corpus (Zhou et al., 2019). In addition, online store's internal search is another source to extract keywords relevant to consumer behaviors (Scholz et al., 2019).

Table 4b. Input/Features for Keyword Generation from Query Logs

| Refs. | Features | | | | | | | | |
|---|---|---|---|---|---|---|---|---|---|
| | SL | URL | CL | SD | KO | HT | SF | DSF | NIS |
| Bartz et al. (2006) | √ | √ | | | | | | | |
| Fuxman et al. (2008) | √ | √ | √ | | | | | | |
| Qiao et al. (2017) | | | | √ | √ | √ | | | |
| Zhou et al. (2019) | | | | | | | √ | √ | |
| Scholz et al. (2019) | | | | | | | | | √ |

Note: SL=Seed Links; URL=Uniform Resource Locator; CL= Clicks; SD=Search Demand (e.g., keyword search demand and joint search demand); KO=Keyword Overlap; HT=Hidden Topic; SF=Semantic Features (e.g., language model score); DSF=Domain-specific Features (e.g., domain category distribution); NIS=The Number of Internal Search Results.

In keyword generation from search result snippets, as we can see from Table 4c, TF and TFIDF, inverse TF, search snippets similarity and common search URLs are taken as predictive variables to characterize the relevance between seed terms and candidate keywords (Wu et al., 2009); and weighted title, meta keywords, meta description and anchor text with its importance inside the HTML document are used to hold semantics of a document (Thomaidou and Vazirgiannis, 2011).

Table 4c. Input/Features for Keyword Generation from Search Result Snippets

| Refs, | Features | | | | | | | |
|---|---|---|---|---|---|---|---|---|
| | IRF | SSF | URL | TI | MF | AT | BT | H1 |
| Wu et al. (2009) | √ | √ | √ | | | | | |
| Thomaidou & Vazirgiannis (2011) | √ | √ | | √ | √ | √ | √ | √ |

Note: IRF=Information Retrieval Oriented Features (e.g., TF and TF-IDF, inverse TF); SSF=Search Snippets related Features (e.g., search snippets similarity, search result snippets); URL=Uniform Resource Locator; TI=Title; MF=Meta related Features (e.g., meta keywords, meta description); AT=Anchor Text; BT=Bold Tags；H1=the Highest Section Level.

In keyword generation from advertising databases, from Table 4d, we can notice that features are relatively scattered. In Chang et al. (2009), feature vectors based on pointwise mutual information are constructed to represent bidding keywords. Similarly, Sarmento et al. (2009) weighted the keyword co-occurrence value feature with mutual information. Schwaighofer et al. (2009) utilized the feature of ad category to suggest keywords based on the prior selected keywords with similar semantics, and Ravi et al. (2010) used keyword overlap, cosine similarity and position of the candidate keyword on the landing page.

Table 4d. Input/Features for Keyword Generation from Advertising Databases

| Reference | Features | | | | | | | | |
|---|---|---|---|---|---|---|---|---|---|
| | SOC | PMI | CO | KO | AC | KO | URL | PK | IRF |
| Chang et al. (2009) | √ | √ | | | | | | | |
| Sarmento et al. (2009) | | | √ | √ | | | | | |
| Schwaighofer et al. (2009) | | | | | √ | | | | |
| Ravi et al. (2010) | | | √ | | | √ | √ | √ | √ |

Note: SOC=Second Order Co-bidding; PMI=Point-wise Mutual Information; CO=Co-occurrence; KO=Keyword Overlap; AC=Ad Category; KO=Keyword Overlap; URL=Uniform Resource Locator; PK=Position of the Keyword (i.e., binary features indicating whether the keyword is present in the title of the landing page, or in its body); IRF=Information Retrieval Oriented Features (e.g., a variant of TF-IDF weighting).

In keyword generation based on semantics and concept hierarchy, semantic relationships between keywords are taken into account for keyword generation (Joshi and Motwani, 2006; Abhishek and Hosanagar, 2007; Mirizzi et al., 2010; Zhang and Qiao, 2018; Zhang et al., 2021), as shown in Table 4e. Similarly, information retrieval oriented features are frequently used to improve the keyword generation results. It is notable to see that Wikipedia is widely used as a corpus to generate keywords for online advertising in various ways, e.g., creating representative vectors for semantic concepts (Amiri et al., 2008; Zhang et al., 2012a) and context vectors (Scaiano and Inkpen, 2011), ranking keyword pairs based on hypertextual links (Mirizzi et al., 2010), mining keywords based on Wikipedia graph (Welch et al., 2010; Jadidinejad and Mahmoudi, 2014).

Table 4e. Input/Features for Keyword Generation based on Semantics and Concept Hierarchy

| Reference | Features | | | | | | | | | | | | | | | | |
|---|---|---|---|---|---|---|---|---|---|---|---|---|---|---|---|---|---|
| | CD | IRF | CO | WF | CH | EIS | HT | URL | TI | HE | AT | HL | MF | SP | OC | AC | LCP | QLF |

| | CD | IRF | CO | WF | CH | EIS | HT | URL | TI | HE | AT | HL | MF | SP | OC | AC | LCP | QLF |
|---|---|---|---|---|---|---|---|---|---|---|---|---|---|---|---|---|---|---|
| Joshi &Motwani (2006) | √ | √ | | | | | | | | | | | | | | | | |
| Abhishek &Hosanagar (2007) | | √ | √ | | | | | | | | | | | | | | | |
| Amiri et al. (2008) | | √ | | √ | | | | | | | | | | | | | | |
| Chen et al. (2008) | | √ | | | √ | | | | | | | | | | | | | |
| Mirizzi et al. (2010) | | | | √ | | √ | | | | | | | | | | | | |
| Welch et al. (2010) | | √ | √ | √ | | | √ | | | | | | | | | | | |
| Scaiano &Inkpen (2011) | | √ | | √ | | | | | | | | | | | | | | |
| Zhang et al. (2012a) | | | | √ | | | | √ | √ | √ | √ | √ | √ | √ | √ | √ | √ | √ |
| Jadidinejad &Mahmoudi (2014) | | | | √ | | | | | | | | | | | | | | |
| Zhang and Qiao (2018) | | | √ | | | | | | | | | | | | | | | √ |
| Nie et al. (2019) | | √ | | √ | | | | | | | | | √ | | | | | |
| Zhang et al. (2021) | √ | √ | | | | | | | | | | | | | | | | |

Note: CD=Characteristic Document (i.e., text-snippets from top 50 search-hits for the keyword); IRF=Information Retrieval Oriented Features (e.g., IDF, TF-IDF, DF, DF-ICF, DF-LCF, CF-IDF); CO=Co-occurrence; WF= Wikipedia Features (e.g., incoming links, outgoing links, redirects, candidate concepts, pool size); CH=Concept Hierarchy; EIS=External Information Sources (e.g., classical search engine results and social tagging); HT=Hidden Topics; URL=Uniform Resource Locator; TI=Title; HE=Headline (i.e., whether it is exactly the headline); AT=Anchor Text; HL=Hyperlink (i.e., whether it is part of a hyperlink of the page); MF=Meta related Features; SP=Span; OC=OneCapt; AC=AllCapt; LCP=Length of the Candidate Phrase; QLF=Query Log Features (e.g., the number of queries related to a keyword in a query log).

### 3.4 Evaluation Metrics for Keyword Generation

In the literature, various evaluation metrics have been employed to evaluate the effectiveness of keyword generation techniques, as summarized in Table 5.

Most research used more than one metric to evaluate the proposed method. Precision, recall, and F-measure are the most commonly used metrics. Among them, precision ranks first, which has been taken into account in 14 out of 30 keyword generation studies; and recall has been considered together with precision in most situations. Precision and recall are regarded as two facets of the quality of keyword generation (Zhou et al., 2007; Berlt et al., 2011). The F-measure family of metrics combines precision and recall, e.g., F1 is the harmonic mean of precision and recall (Zhou et al., 2019). Li et al. (2007) used the F1 from both macro (i.e., the transcript level) and micro (i.e., the section level) perspectives.

Besides these three popular metrics, a rich set of metrics has been used to measure whether the generated keyword set is satisfactory, useful or effective, such as relevance (Thomaidou and Vazirgiannis, 2011), novelty (Qiao et al., 2017; Zhou et al., 2019; Zhang et al., 2021) and coverage (Chang et al., 2009; Nie et al., 2019). Joshi and Motwani (2006) developed a variant of F-measure taking into account maximizing precision, recall and nonobviousness to measure the goodness of keyword generation. Because the ultimate goal of keyword generation is to improve advertising performance and bring more profits to advertisers, performance indexes such as click-through rate are also considered as important metrics (Sarmento et al., 2009; Zhou et al., 2019).

Table 5. Evaluation Metrics for Keyword Generation

| **Metric** | **Definition** | **Refs.** |
|---|---|---|
| Top-n score | The number of the top-n outputs that are in the list of terms described by the annotator for that page, e.g., Top-10 score. | Yih et al. (2006); Lee et al. (2009) |
| Entropy | Let $P(k|p)$ denote the probability that keyword $k$ is relevant to page $p$, $Entropy = -\log_2 P(k|p)$ if $k$ is relevant to $p$; $Entropy = -\log_2(1 - P(k|p))$, otherwise. | Yih et al. (2006) |
| Accuracy | The ratio of correctly predicted observations to the total observations. $Accuracy = (True\ postitive + True\ negative)/(True\ postitive + True\ negative + False\ postitive + False\ negative)$. | Berlt et al. (2011); Zhou et al. (2019) |
| Precision | The ratio of correctly predicted positive observations to the total predicted positive observation. | Wu &Bolivar (2008); Berlt et al. (2011); Zhou et al. (2007); Li et al. (2007); Bartz et al. (2006); Zhou et al. |

|  | $Precision = True\ positive/$ $(True\ positive + False\ positive)$. | (2019); Wu et al. (2009); Chang et al. (2009); Joshi &Motwani (2006); Abhishek &Hosanagar (2007); Chen et al. (2008); Mirizzi et al. (2010); Zhang et al. (2012a); Zhang et al. (2021) |
|---|---|---|
| Recall | The ratio of correctly predicted positive observations to all observations in actual class. $Recall = True\ positive/(True\ positve + False\ negaitve)$. | Berlt et al. (2011); Zhou et al. (2007); Li et al. (2007); Zhou et al. (2019); Joshi &Motwani (2006); Abhishek &Hosanagar (2007); Chen et al. (2008); Zhang et al. (2021) |
| F1-measure | The weighted average of Precision and Recall. $F1 = 2 * Precision * Recall/(Precision + Recall)$. | Li et al. (2007); Qiao et al. (2017); Zhou et al. (2019); Joshi &Motwani (2006); Chen et al. (2008); Zhang &Qiao (2018); Zhang et al. (2021) |
| NDCG | The normalized discounted cumulative gain, i.e., a measure of ranking quality. It is calculated for the sorted list of results for each of the keywords $NDCG = N \sum_{i=1}^{k} \frac{2^{score(j)}-1}{\log_2(j+1)}$, where N is the normalization constant chosen so that a perfect ordering of the results will receive the score of one; $score(j)$ is the gain value associated with the label of the item at the $j$-th position of the ranked list; $\log_2(j + 1)$ is a discounting function that reduces the document's gain value as its rank increases. | GM et al. (2011); Chen et al. (2008) |
| Percentage of URL Match | A measure of the efficiency of a hierarchy in matching unindexed URLs which is defined as the percentage of correctly matched URL in a given hierarchy. $Percentage\ URL\ Match = \frac{\sum_{(k,v) \in u} 1_{(k,v) matched hierarchy}}{|u|}$, where $(k, v)$ is a key-value pair for an URL, and $|u|$ is the set of all key-value pairs for an URL. | GM et al. (2011) |
| Jaccard Similarity | A measure of similarity between two keyword sets which is defined as the size of the intersection divided by the size of the union of the sample sets. $Jaccard\ Similarity = \frac{h_{feats} \cap s_{feats}}{h_{feats} \cup s_{feats}}$, | GM et al. (2011) |

| | where $s_{feats}$ and $h_{feats}$ are keywords and categories from semantic and matched path in the hierarchy. | |
|---|---|---|
| Relevance | The fraction of relevant keywords in the set. | Fuxman et al. (2008); Thomaidou &Vazirgiannis (2011); Joshi &Motwani (2006); Mirizzi et al. (2010); Nie et al. (2019) |
| Indirectness | The fraction of keywords indirectly connected to the seed set in the result set, a query $q$ is indirect if $q \notin Q(S)$, where $Q(S)$ denotes the query set directly connected to seed set $S$. | Fuxman et al. (2008) |
| Novelty | The fraction of new effective keywords which are omitted or not found by other methods. | Qiao et al. (2017); Zhou et al. (2019); Zhang &Qiao (2018); Zhang et al. (2021) |
| Keyword number | The number of generated keywords. | Scholz et al. (2019) |
| Impression | The number of times that an advertisement is displayed on results pages. | Scholz et al. (2019); Scaiano &Inkpen (2011) |
| Cost per click | Advertiser's pay for each click on the ads. | Scholz et al. (2019) |
| Conversion rate | The number of conversions divides the number of total ad interactions that can be tracked to a conversion during the same period. | Scholz et al. (2019) |
| Perplexity | A measurement of how well a probability distribution or probability model predicts a sample, specifically the generation quality with respect to grammar and fluency computed by the generation distribution in the models. $Perplexity(p) = 2^{H(p)}$, where $H(p)$ is the entropy of the distribution. | Zhou et al. (2019) |
| Distinct-n | The proportion of distinct n-grams to all the n-grams in generated keywords to evaluate the diversity. | Zhou et al. (2019) |
| Coverage | The ratio between the number of suggestions produced by a system and the maximum number of allowed suggestions. | Zhou et al. (2019); Chang et al. (2009); Nie et al. (2019) |
| Click-through rate | The ratio of page views that lead to a click to the total number of page views. | Zhou et al. (2019); Sarmento et al. (2009) |
| Revenue per mille | The revenue of a search engine per one thousand page views. | Zhou et al. (2019) |
| Specificity | How general or specific were the generated keywords judged by some researchers and students. | Thomaidou &Vazirgiannis (2011) |
| Nonobviousness | A term not containing the seed keyword or its variants sharing a common stem. | Thomaidou &Vazirgiannis (2011); Joshi &Motwani (2006) |

| Automatic suggestion ratio | The ratio between the number of automatically suggested keywords and the number of user-suggested keywords associated with the ad. $S_r(i) = \frac{f_{auto}(i)}{f_{user}(i) + f_{auto}(i)}$, where $f_{auto}(i)$ and $f_{user}(i)$ are the number of automatically suggested keywords and the number of user-suggested keywords associated with the $i$-th ad. | Sarmento et al. (2009) |
|---|---|---|
| Average rank | The average rank of the suggested keywords. | Sarmento et al. (2009) |
| Keyword ranking index | The fraction of suggested keywords at rank n (e.g., rank 1 or rank 10) selected by the user. | Sarmento et al. (2009) |
| Average keyword printability | The average number of ad prints (i.e., impression) that are made as result of a bid placed on a suggested keyword. | Sarmento et al. (2009) |
| Average ad printability | The average number of times an ad is printed as result of an automatically suggested keyword. | Sarmento et al. (2009) |
| Average keyword clickability | The average number of clicks made on ads that are printed as result of a bid placed on a suggested keyword. | Sarmento et al. (2009) |
| Average ad clickability | The average number of times an ad is clicked as a result of an automatically suggested tag. | Sarmento et al. (2009) |
| Keyword printability efficiency | The fraction of suggested keywords that lead the corresponding ads being printed (i.e., displayed). | Sarmento et al. (2009) |
| Keyword clickability efficiency | The fraction of suggested keywords that lead the corresponding ads being clicked. | Sarmento et al. (2009) |
| True/False positive ratio | The fraction of the pairs correctly/incorrectly classified as belonging to the same cluster. | Schwaighofer et al. (2009) |
| Negative log likelihood of the test set | Closely related to the log perplexity, a quality criterion that has been used to evaluate for the LDA model. | Schwaighofer et al. (2009) |
| Advertiser entropy scores | The entropy of the distribution of advertisers across clusters. | Schwaighofer et al. (2009) |
| Edit distance | The average number of words required to be inserted, deleted or substituted within the phrase $k$ in order to convert it to the gold phrase $k^*$. $ED(k, k^*) = \frac{\# \text{ of operations to convert } k \rightarrow k^*}{\# \text{ of words in } k^*}$. | Ravi et al. (2010) |

| Rouge-1 score | A recall-based measure which evaluates the quality of a candidate bid phrase against all the relevant gold bid phrases. $ROUGE - 1(k, l) = \frac{\sum_{k^* \in K^*} \# \ of \ words \ in \ k \cap k^*}{\sum_{k^* \in K^*} \# \ of \ words \ in \ k^*}$, where $l$ is the test landing page. | Ravi et al. (2010) |
|---|---|---|
| Ad display | Whether campaign ads (or very closely related ads) are shown or not. | Scaiano &Inkpen (2011) |
| Popularity | An indicator of how pertinent the keywords are to advertisers. $Popularity = \frac{1}{|R(S)|}\sum_{k \in R(S)} A_k^*$, where $R(S)$ are the keywords from source $S$ judged relevant by at least one user, and $A_k^*$ is the number of advertisers bidding for keyword $k$. | Welch et al. (2010) |

### 3.5 Summary

In summary, all the five sources of information are valuable to build domain-specific keyword pools. In this sense, none of keyword generation methods based on a single information source can provide the perfect solution because each of them has its own advantages and shortcomings. We believe that the five branches of works complement each other, and thus it calls for a benchmark study integrating the five sources of information to generate relevant keywords.

## 4. Keyword Targeting

### 4.1 Problem Description

In SSA, advertisers need to select a specific set of keywords from the domain-specific keyword pool and determine appropriate match types (i.e., broad match, phrase match and exact match) for these keywords in order to reach a specific target population. Based on selected keywords, their advertisements are displayed when users submit queries or browse Web pages. We term this process as keyword targeting following the marketing paradigm (analog to the targeting strategy), or market-level keyword optimization from the perspective of decision making. Note that keyword targeting has a more broad sense than keyword selection. The output of keyword targeting is a set of keywords and corresponding match types called the target keyword set.

Formally, given that a set of generated keywords $K^{(GNT)}$ is determined, keyword targeting can be defined as follows.

$$f^{(TGT)}: K^{(GNT)} \xrightarrow{x_{i,\bar{m}}^{(TGT)}} K^{(TGT)}, K^{(TGT)} \subseteq K^{(GNT)},$$

$$\sum_{\bar{m}=0}^{3} x_{i,\bar{m}}^{(TGT)} = 1,$$

$$x_{i,\bar{m}}^{(TGT)} = \begin{cases} 1, & \text{if } k_i \text{ is selected in match type } \bar{m} \\ 0, & \text{otherwise} \end{cases},$$

$$i \in \{1,2,\ldots,n^{(GNT)}\}, \bar{m} \in \{0,1,2,3\}, \qquad (26)$$

where $f^{(TGT)}$ is the keyword targeting function, $K^{(TGT)}$ is a set of selected keywords from the domain-specific keyword pool ($K^{(GNT)}$), and $x_{i,\bar{m}}^{(TGT)}$ is the decision variable of keyword targeting, indicating whether the $i$-th keyword is selected in match type $\bar{m}$. Note that each keyword can be selected in one and only one match type or is not selected at all. Hence, $\sum_{\bar{m}=0}^{3} x_{i,\bar{m}}^{(TGT)} = 1$, $\bar{m} = 0,1,2,3$ indicate null, exact match, phrase match and broad match, respectively.

In SSA, many keywords do not raise reliable advertising impacts, and instead occupy a large portion of the advertising expenditure. Keyword targeting can prevent advertisers from targeting wrong populations that waste their advertising resources with poor returns. Moreover, it's critical to select keywords because consumers with different intents tend to use different types of keywords (e.g., general keywords and specific keywords). Thus, product sales via SSA highly rely on an effective set of keywords that describe advertised offerings and consumers' intents.

However, it's not a straightforward task to select a set of right keywords from a large pool with millions of available keywords (Bartz et al., 2006). Even if the size of the keyword pool is relatively small, it's unwise to bid on all keywords simultaneously (Ji et al., 2010). Thus, keyword targeting is one of the most crucial steps in search advertising optimization. In effect, keyword targeting is a cornerstone process for all SSA stakeholders (i.e., search engines, advertisers, and consumers) pursuing the best combination of advertising presentation, promotion, and discovery (Thomaidou and Vazirgiannis, 2011). Moreover, the keyword targeting process needs to take into account mechanisms behind search engines, product/service features, as well as characteristics of the target population.

There are two tradeoffs in the keyword targeting process (Rusmevichientong and Williamson, 2006). First, there is a tradeoff between selecting a limited number of profitable keywords versus selecting an extensive set of keywords (i.e., the more-less tradeoff). The former will not spend the entire budget, while the latter will deplete the budget quickly, in turn which might lose opportunities to receive clicks and conversions that may arrive later. Second, advertisers must balance the tradeoff between selecting known keywords that yielded high profits in the past versus selecting previously unused keywords whose performance indexes such as click-through probabilities can be learnt in future advertising campaigns (i.e., the exploitation-exploration tradeoff)[1].

In the literature, keyword targeting has been treated as two independent problems, namely keyword selection and keyword match. In the following, we discuss these two research streams in terms of techniques, features and evaluation metrics.

**4.2 Techniques for Keyword Targeting**

**4.2.1 Keyword Selection**

Given a certain amount of budget, advertisers try to spend their money on the most profitable keywords. In a highly uncertain environment such as SSA (Yang et al., 2013; Li and Yang, 2020), identifying a set of profitable set of keywords becomes even more challenging for advertisers (Yang et al., 2019).

In the literature, techniques used for keyword selection, include feature selection, adaptive approximation, mixed integer optimization, technique for order of preference by similarity to ideal solution (TOPSIS) and the mean-variance model, as summarized in Table 6a. Table 7a presents notations used in keyword selection.

Table 6a. Techniques for Keyword Selection

| Category | Approach | Refs. |
| --- | --- | --- |
| Feature selection | Information gain | Kiritchenko &Jiline (2008) |
| | Symmetrical uncertainty | Kiritchenko &Jiline (2008) |
| | Chi-square statistics | Kiritchenko &Jiline (2008) |
| | Odds ratio | Kiritchenko &Jiline (2008) |
| | Precision on the positive class | Kiritchenko &Jiline (2008) |
| Optimization | Adaptive approximation | Rusmevichientong &Williamson (2006) |
| | Mixed integer optimization | Zhang et al. (2014b) |
| | Technique for order of preference by similarity to ideal solution | Arroyo-Cañada &Gil-Lafuente (2019) |
| | Mean-variance model | Symitsi et al. (2022) |

---

[1] In the reinforcement learning literature, it is cast as balancing the exploitation of known good options and the exploration of unknown options that might be better than known options.

Table 7a. Notations in keyword selection

| Terms | Definition |
|---|---|
| $n_k$ | The number of query keywords |
| $\_k_i$ | The absence of $k_i$ |
| $\mathbb{C}$ | A category set |
| $\_\mathbb{C}_j$ | The categories in $\mathbb{C}$ other than $\mathbb{c}_j$ |
| $p_i$ | The cost-per-click for keyword $k_i$ |
| $c_i$ | The click-through rate for keyword $k_i$ |
| $z_i$ | The expected profit from keyword $k_i$ |
| $\lambda_i$ | The probability that keyword $k_i$ is queried |
| $\delta^a$ | The upper bound of the expected number of queries for keyword $k_i$ |
| $u$ | The mean of the total number of queries that arrive in a period |
| $\ell$ | The largest prefix for keyword selection |
| $\mathfrak{i}_t$ | An integer random variable in period $t$ |
| $\mathfrak{b}_t$ | An independent binary random variable in period $t$ |
| $\gamma_t$ | The probability that $\mathfrak{b}_t = 0$, i.e., $\gamma_t = P\{\mathfrak{b}_t = 0\}$ |
| $\mathfrak{g}_t$ | The largest index for keyword selection in period $t$ |
| $d_i^t$ | The number of impressions that keyword $k_i$ receives in period $t$ |
| $\mathfrak{c}_i^t$ | The number of clicks that keyword $k_i$ receives in period $t$ |
| $sum\_d_i$ | The cumulative number of impressions for keyword $k_i$ in periods from 1 to $t$ |
| $sum\_\mathfrak{c}_i$ | The cumulative number of clicks for keyword $k_i$ in periods from 1 to $t$ |
| $\hat{c}_i^{t-1}$ | The estimated click-through rate (CTR) for keyword $k_i$ in period $t$ |
| $\Theta_i$ | The auctions triggered by keyword $k_i$ |
| $\vartheta$ | An auction |
| $K_l$ | A set of keyword for the $l$-th ad-group |
| $v_l$ | The average true value of a click for the $l$-th ad-group |
| $b_{l,i}$ | The bid price for keyword $k_i$ in the $l$-th ad-group |
| $d_{l,\vartheta}$ | The impression probability for the $l$-th ad-group in auction $\vartheta$ |
| $c_l$ | The average click-through rate for the $l$-th ad-group |
| $b_\epsilon$ | The floor bid price |
| $\omega_\vartheta$ | The winning score of auction $\vartheta$ |
| $r_{l,\vartheta}$ | The relevance score of the $l$-th ad-group in auction $\vartheta$ |
| $z_{ij}$ | The performance value of the $i$-th set of keywords under the $j$-th evaluation criterion |
| $w_j$ | The weight of the $j$-th criterion |
| $v_{ij}$ | The weighted normalized value of the $i$-th set of keywords under the $j$-th evaluation criterion |
| $A^+$ | The positive ideal solution |
| $A^-$ | The negative ideal solution |
| $J$ | A benefit criteria index |
| $J'$ | A cost criteria index |
| $a_i^+$ | An alternative from the positive ideal solution $A^+$ |
| $a_i^-$ | An alternative from the negative ideal solution $A^-$ |
| $R_i$ | The relative closeness |

| $u_{A_+}(x_j), u_{A_-}(x_j)$ | The characteristic membership functions |
| --- | --- |
| $w_i$ | The percentage allocation of the budget for the $i$-th keyword |
| $z_i$ | The growth in profit for the $i$-th keyword |

**(1) Feature selection (FS).** By assuming that keywords can be optimized based on their historic performance, Kiritchenko & Jiline (2008) applied a set of feature selection techniques to a set of words (i.e., search terms) combinations (i.e., multi-word phrases) comprising historical users' queries to optimize keyword selection. More specifically, the past performance of individual keywords and all possible multi-word keywords is analyzed, then feature selection techniques were used to sort phrases according to their effectiveness extracted from the historical data, and then a set of profitable phrases was selected. Let $n_k$ denote the number of query keywords (i.e., single words and word combinations), $k_i$ denote a sequence of non-space characters, $\_k_i$ represent the absence of $k_i$, and $\_\mathbb{c}_j$ denote categories in $\mathbb{C}$ other than $\mathbb{c}_j$. The relevance of $k_i$ to category $\mathbb{c}_j \in \mathbb{C}$ can be measured by a function such as information gain, chi-square statistic, symmetrical uncertainty, odds ratio, and precision on the positive class, which are described as follows.

**Information gain** is the amount of information gained about a random variable, which can tell how important a given attribute is in a feature vector. The information gain-based relevance of $k_i$ to category $\mathbb{c}_j$ is given as $H(\mathbb{C}) - H(\mathbb{C}|A)$, where $H(\mathbb{C}) = -\sum_j P(\mathbb{c}_j) \log_2 P(\mathbb{c}_j)$, and $H(\mathbb{C}|A) = -\sum_{A \in \{k_i, \_k_i\}} P(A) \sum_j P(\mathbb{c}_j|A) \log_2 P(\mathbb{c}_j|A)$.

**Chi-square statistic** is a measure of the difference between observed and expected frequencies of outcomes of a set of events or variables. The Chi-square-based relevance of $k_i$ to category $\mathbb{c}_j$ is given as $\frac{n_k(P(k_i,\mathbb{c}_j)P(\_k_i,\_\mathbb{c}_j) - P(k_i,\_\mathbb{c}_j)P(\_k_i,\mathbb{c}_j))^2}{P(k_i)P(\_k_i)P(\mathbb{c}_j)P(\_\mathbb{c}_j)}$, where $P(k_i,\mathbb{c}_j)P(\_k_i,\_\mathbb{c}_j)$ is the probability that keywords containing $k_i$ is related to $\mathbb{c}_j$ and keywords excluding $k_i$ is related to categories other than $\mathbb{c}_j$; $P(k_i,\_\mathbb{c}_j)P(\_k_i,\mathbb{c}_j)$ is the probability that keywords containing $k_i$ is related to categories other than $\mathbb{c}_j$ and keywords excluding $k_i$ is related to $\mathbb{c}_j$.

**Symmetrical uncertainty** measures the relevance between a feature and the class label. The symmetrical uncertainty-based relevance of $k_i$ to category $\mathbb{c}_j$ is given as $2\left[\frac{H(\mathbb{C}) - H(\mathbb{C}|A)}{H(\mathbb{C}) + H(A)}\right]$.

**Odds ratio** is a measure of association between an exposure and an outcome, which represents the odds that an outcome will occur given a particular exposure, compared to the odds of the outcome occurring in the absence of that exposure. The odds ratio-based relevance of $k_i$ to category $\mathbb{c}_j$ is given as $\frac{P(k_i|\mathbb{c}_j)(1-P(k_i|\_\mathbb{c}_j))}{(1-P(k_i|\mathbb{c}_j))P(k_i|\_\mathbb{c}_j)}$.

**Precision on the positive class** measures the fraction of keywords containing word $k_i$ which is relevant to category $\mathbb{c}_j$, i.e., $P(\mathbb{c}_j|k_i)$.

Feature selection techniques discussed above showed similar performance. Among them, symmetrical uncertainty performed the best by a slight margin and the precision on the positive class technique is a little inferior to others. In general, feature selection techniques could not only identify profitable keywords, but also discover more specific phrases.

**(2) Adaptive approximation (AA)**. In light of the more-less tradeoff and the exploitation-exploration tradeoff, Rusmevichientong and Williamson (2006) developed adaptive approximation algorithms to solve the keyword selection problem.

In the static case where click-through rates of keywords are known, keyword selection can be modeled as a stochastic knapsack problem with query arrival as a random variable. Let $p_i$, $c_i$, $z_i$ denote the cost-per-click, click-through rate and expected profit for keyword $k_i$, respectively. Let $\lambda_i$ denote the probability that $k_i$ is queried and $u$ denote the mean of the total number of queries that arrive in a period. In order to develop an efficient approximation algorithm for keyword selection, keywords are sorted in a prefix-orderings, i.e., the descending order of profit-to-cost ratio, i.e., $\frac{z_1}{p_1} \geq \frac{z_2}{p_2} \ldots \geq \frac{z_N}{p_N}$ and the expected number of queries for $k_i$ (i.e., $\lambda_i u$) is at most $\delta^{\mathfrak{a}}$ ($0 \leq \mathfrak{a} < 1, \delta \geq 1$, and $p_i \leq \frac{1}{\delta}$). Then a near-optimal approximation algorithm was developed to choose the largest prefix $\ell$ which satisfies

$$max\left\{\ell: u\sum_{i=1}^{\ell} p_i c_i \lambda_i \leq 1 - \frac{1}{\delta} - \frac{1}{\delta^{(1-\mathfrak{a})/3}}\right\}. \qquad (27)$$

That is, the near-optimal selected set includes keywords with prefix-orderings $\{1,2,\ldots,\ell\}$, whose cost is close to the budget.

In the dynamic case where the click-through rates are not known, keyword selection was formulated as a multi-armed bandit problem. An improved adaptive approximation algorithm was developed to select a bandit (i.e., a subset of keywords) in each time period based on their past observations that yield near-optimal profits. In this algorithm, the click-through rates for $k_i$ are updated according to impressions $d_i^t$ and clicks $\mathfrak{c}_i^t$ that the keyword receives in period $t$: as for all $i$, $sum\_d_i := sum\_d_i + d_i^t$, $sum\_c_i := sum\_c_i + c_i^t$, and estimated click-through rate $\hat{c}_i^{t-1} = \frac{sum\_c_i}{sum\_d_i}$, if $sum_{d_i} > 0$; $\hat{c}_i^{t-1} = 1$, otherwise. Let $\mathfrak{i}_t$ denote an integer randomly chosen uniformly from the set $\{1,2,\ldots,n_k\}$ in period $t$ and $\mathfrak{b}_t$ denote an independent binary random variable with $P\{\mathfrak{b}_t = 1\} = 1 - \gamma_t$ and $P\{\mathfrak{b}_t = 0\} = \gamma_t$, where $\gamma_t \in [0,1]$. Similarly, keywords were sorted in the descending order of profit-to-cost ratio. Keywords with indexes $\{1,2,\ldots,\mathcal{g}_t\}$ are selected as their target set, where $\mathcal{g}_t = \ell_t$, if $\mathfrak{b}_t = 1$; $\mathcal{g}_t = \mathfrak{i}_t$, otherwise; and $\ell_t$ is the index such that

$$u \sum_{i=1}^{\ell_t} p_i \hat{c}_i^{t-1} \lambda_i \leq 1 - \frac{1}{k} - \frac{2}{k^{(1-\alpha)/3}} \leq u \sum_{i=1}^{\ell_t+1} p_i \hat{c}_i^{t-1} \lambda_i. \qquad (28)$$

The AA algorithm outperformed multi-armed bandit algorithms by increasing about 7% profits, and the expected profit could converge to a near-optimal level.

**(3) Mixed integer optimization (MIP)**. MIP adds an additional condition that at least one of the variables can only take integer values on the basis of linear programming which maximizes (or minimizes) a linear objective function subject to one or more constraints. In the SSA context, Zhang et al. (2014b) modeled keyword selection as a MIP problem, which maximizes an advertiser's revenue and the relevance of selected keywords, while minimizing the keyword competitiveness, with constraints of the lower and the upper bounds of bidding prices on a set of keywords and the limited budget for an ad-group.

Given an ad-group $l$, a keyword set for $l$ (i.e., $K_l$) can be obtained by filtering out keywords whose relevance scores are less than a certain threshold. Let $\Theta_i$ denotes auctions triggered by keyword $k_i$. In an auction $\vartheta \in \Theta_i$, the expected revenue of ad-group $l$ from $k_i$ is $\sum_{\vartheta \in \Theta_i}(v_l - b_{l,i})d_{l,\vartheta}c_l$, where $v_l$ is the average true value of a click, $b_{l,i}$ is the bidding price, $d_{l,\vartheta}$ is the impression probability, and $c_l$ is the average CTR. Let $x_i^{(SEL)} = 1$ denote the indicator variable for keyword selection if $k_i$ is selected; otherwise $x_i^{(SEL)} = 0$. In order to maximize the total expected revenue, the objective is given as

$$\max_{x_i^{(SEL)}, b_{l,i}} \sum_{k_i \in K_l} \left\{ c_i \sum_{\vartheta \in \Theta_i}(v_l - b_{l,i})d_{l,\vartheta} \, c_l x_i^{(SEL)} \right\}, \qquad (29)$$

where $c_i$ is the impression confidence based on keyword competitiveness.

Constraints are the budget of an ad-group, and the lower bound and the upper bound of bidding prices, which are given as

$$s.t. \sum_{k_i \in K_l} \left\{ \sum_{\vartheta \in \Theta_i} b_{l,i} d_{l,\theta} \, c_l x_i^{(SEL)} \right\} \leq B_l, \text{ with } \max_{\vartheta \in \Theta_i} \left\{ b_\epsilon, \frac{\omega_\vartheta}{r_{l,\vartheta}} \right\} \leq b_{l,i} \leq v_l, \qquad (30)$$

where $b_\epsilon$ is the floor price, $\omega_\vartheta$ is the winning score of auction $\vartheta$, and $r_{l,\vartheta}$ is the relevance score of ad-group $l$ in $\vartheta$.

In order to select relevant yet less-competitive keywords and put optimal bidding prices over these keywords, Zhang et al. (2014b) constructed a mixed integer programming model and solved it by iteratively conducting binary integer programming and sequential quadratic programming until convergence. Simulation experiments showed that the MIP-based keyword selection method is capable of increasing impressions, expected clicks, advertiser's revenue, as well as search engine's revenue.

**(4) Technique for order of preference by similarity to ideal solution (TOPSIS).** TOPSIS is a multi-criteria decision analysis method based on the concept that the chosen alternative should have the shortest distance from the positive ideal solution (PIS) and the furthest distance from the negative ideal solution (NIS) (García-Cascales and Lamata, 2012). Given several alternative sets of keywords for an advertising campaign, let $z_{ij}$ denote the performance value of the $i$-th set of keywords under the $j$-th evaluation criterion $z_j$ (e.g., advertising inventory, impressions per week, clicks per week, opportunity to see, click-through rate, cost per click, and revenue), $i = 1,2,\ldots,n$, $j = 1,2,\ldots,m$, and $w_j$ denote the weight of the $j$-th criterion. The weighted normalized value of the $i$-th set of keywords under the $j$-th evaluation criterion is calculated as

$$v_{ij} = w_j \frac{z_{ij}}{\sqrt{\sum_{j=1}^n (z_{ij})^2}}. \qquad (31)$$

The positive ideal value set $A^+$ and the negative ideal solution $A^-$ are determined as

$$A^+ = \{v_1^+, \ldots, v_m^+\} = \left\{ (\max_i v_{ij}, j \in J)(\min_i v_{ij}, j \in J') \right\}, \text{ and}$$

$$A^- = \{v_1^-, \ldots, v_m^-\} = \left\{ (\min_i v_{ij}, j \in J)(\max_i v_{ij}, j \in J') \right\}, \qquad (32)$$

where $J$ is associated with benefit criteria, and $J'$ is associated with cost criteria.

Then the separation of an alternative from the positive ideal solution (PIS) $A^+$ and the negative ideal solution (NIS) $A^-$ is given as

$$a_i^+ = \sum_{j=1}^m |v_{ij} - v_j^+|, \text{ and}$$

$$a_i^- = \sum_{j=1}^m |v_{ij} - v_j^-|. \qquad (33)$$

Then the relative closeness $R_i$ to the ideal solution can be expressed as $R_i = \frac{a_i^-}{a_i^+ + a_i^-}$, and keywords can be selected according to rank $R_i$ in descending order. Arroyo-Cañada & Gil-Lafuente (2019) proposed a fuzzy asymmetric TOPSIS-based keyword selection method. Specifically, this method introduced fuzzy indicators by replacing $z_{ij}$ with fuzzy numbers or linguistic values, and incorporated characteristic membership functions $u_{A_+}(z_j)$ and $u_{A_-}(z_j)$ to asymmetrically penalize the lack of frequency and soften light excesses for the $j$-th criterion, i.e., $R_i = \sum_{j=1}^{n}|v_{ij} - v_j^-|^{u_{A_-}(z_j)}/(\sum_{j=1}^{n}|v_{ij} - v_j^+|^{u_{A_+}(z_j)} + \sum_{j=1}^{n}|v_{ij} - v_j^-|^{u_{A_-}(z_j)})$.

**(5) Mean-variance model (MVM)**. MVM selects the most efficient portfolio by analyzing expected returns (mean) and standard deviations (variance) of various portfolios (Markowitz, 1952). The efficient frontier of keyword portfolios can be obtained as follows.

$$\min_{w_i} \sum_{i=1}^{n_k}\sum_{j=1}^{n_k} w_i w_j cov(z_i, z_j),$$

$$s.t., \mu_p = \sum_{i=1}^{n_k} w_i E(z_i), \sum_{i=1}^{n_k} w_i = 1, w_i \geq 0, \quad (34)$$

where $w_i$ and $z_i$ are the percentage allocation of the budget and the growth in profit for the $i$-th keyword, respectively.

Symitsi et al. (2022) explored keyword portfolios by examining the risk-adjusted keyword performance, and selected unrelated and negatively related keywords into keyword portfolios for the goal of diversification. The MVM-based keyword selection method outperformed advertisers' heuristic rules used in practice.

In summary, current keyword selection methods can recognize profitable keywords under budget constraints by balancing the tradeoff between costs and revenues. However, the dynamic feature of SSA has been ignored in prior studies on keyword selection, which assumes that keyword costs and revenues are unchangeable (Rusmevichientong and Williamson, 2006; Kiritchenko and Jiline, 2008).

**4.2.2 Keyword Match**

In SSA, advertisers need to make another important choice over keyword match types, including exact match, phrase match and broach match, when making keyword targeting decisions (Dhar and Ghose, 2010). Keyword match type controls when advertisements will be shown to consumers, which in turn determines the target population of potential consumers. Therefore, keyword match is a critical variable in SSA (Du et al., 2017; Yang et al., 2021).

In the literature on keyword match, current research primarily focused on broad match, which falls into two research streams. One stream explored broad match mapping mechanisms to help advertisers identify similar keywords, increase the advertising reach and reduce the campaign management burden, by using regression SVM, max-margin voted perceptron and distributed language model (Radlinski et al., 2008; Gupta et al., 2009; Grbovic et al., 2016). Another stream addressed optimization problems in broad match, using graph model and game-theoretic model (Singh and Roychowdhury, 2008; Amaldoss et al., 2016). Techniques used for keyword match in the literature are summarized in Table 6b. Table 7b presents notations used in keyword match.

Table 6b. Techniques for Keyword Match

| Category | Approach | Refs. |
| --- | --- | --- |
| Learning model | Support vector regression | Radlinski et al. (2008) |
| | Max-margin voted perceptron | Gupta et al. (2009) |
| | Distributed language model | Grbovic et al. (2016) |
| Graph model | / | Singh &Roychowdhury (2008) |
| Game theory | Game-theoretic model | Amaldoss et al. (2016) |

Table 7b. Notations in keyword match

| Terms | Definition |
| --- | --- |
| $x_i$ | The $i$-th training sample |
| $y_i$ | The target value of the $i$-th training sample |
| $\tau$ | The intercept of a prediction |
| $ð$ | A threshold |
| $x = f(k \to k')$ | A feature vector encoding various properties of an advertisement's impressions on a broad match keyword $k'$ shown in a context containing keyword $k$ |
| $c$ | A binary variable indicating whether keyword $k'$ is clicked |
| $\hat{c}$ | The click probability of a broad match keyword |
| $w$ | A weight vector |
| $n_{w_i}$ | The number of iterations that the $i$-th weight vector don't change |
| $w_{avg}$ | A moving average weight vector |
| $\zeta$ | An amnesia rate |
| $\hat{S}$ | A set of search sessions |
| $s$ | A search session $s = (a_1, ..., a_N) \in \hat{S}$ |
| $v_{a_n}$ | A multiple dimensional real-valued representation for a unique action $a_n$ |
| $\xi$ | The length of the relevant context for action sequences |
| $\Phi$ | A vocabulary set for unique actions in the dataset |
| $n_K$ | The number of keywords in set $K$ |
| $n_{Ader}$ | The number of advertisers who are interested in the keywords |
| $n_{AS}$ | The maximum number of ad-slot available |
| $\mathcal{E}$ | The valuation matrix with entries $\varepsilon_{j,k}$ |
| $\varepsilon_{j,k}$ | The product of true value and quality score of advertiser $ader_j$ for keyword $k$ |

| $B_j$ | The daily budget for advertiser $ader_j$ |
| --- | --- |
| $d_k$ | The daily search demands of keyword $k$ |
| $Ader$ | A set of advertisers |
| $S$ | A set of edges connecting advertisers and keywords in broad match graph |
| $S'$ | The extension set of edges, $S \subset S'$ |
| $G$ | A graph $G = (Ader, K, S)$ |
| $G'$ | A broad-match graph for graph $G$ |
| $EU$ | The expected utility |

**(1) Support vector regression (SVR)**. SVR is a supervised learning model for regression analysis (Scholkopf, 1999). Given a training sample $x_i$ with a target value $y_i$ and the prediction of the inner-product plus intercept $\langle w, x_i \rangle + \tau$ for that sample, SVR is defined as

$$\min \frac{1}{2} \|w\|^2$$

$$s.t. |y_i - \langle w, x_i \rangle - \tau| \leq ð, \quad (35)$$

where ð is a threshold.

Radlinski et al. (2008) presented a two-stage method combining exact match and broad match to recommend advertisements, with the objective of optimizing both the advertising relevance and the advertising revenue for search engines simultaneously. At the first stage, an ad query substitution table was built using external knowledge sources in an offline setting. Specifically, they fixed a large set of sufficiently frequent ad queries, and used SVR to learn weights for features in a combined linear function computing the final score for each query substitute. At the second stage, advertisements retrieval was performed by finding advertisements whose keywords exactly match the substituted query. The SVR-based method combines merits of both broad match (i.e., flexibility) and exact match (i.e., computational efficiency).

**(2) Max-margin voted perceptron (MMVP)**. MMVP is a discriminative online classifier performing well on high-dimensional learning tasks (Freund & Schapire, 1999). Assume that for every keyword ($k$), there exists a set of broad-match keywords ($\{k'\}$) that can be identified using some similarity functions. The problem of identifying broad match for a given keyword is equal to predicting the click probability of broad match keyword. Let $x = f(k \rightarrow k')$ denote a feature vector encoding various properties of the impression of an advertisement on a broad match keyword $k'$ shown in a context containing keyword $k$, and $c \in \{-1, +1\}$ denotes a binary variable indicating whether $k'$ is clicked. The dataset of training instances for MMVP in SSA is constructed as $\{(x = f(k \rightarrow$ $k'), c)\}, c \in \{-1, +1\}$. The MMVP algorithm starts with an initial zero weight vector $\boldsymbol{w} = \boldsymbol{0}$, and predicts the click probability of a broad match keyword, i.e., $\hat{c} = sigmoid(\boldsymbol{w} \cdot \boldsymbol{x})$. If the prediction $\hat{c}$ is different from the label $c$, it updates the weight vector $\boldsymbol{w} := \boldsymbol{w} + c\boldsymbol{x}$, otherwise $\boldsymbol{w}$ is not changed. The process runs repeatedly through all the training instances. In the process, the number of iterations that each weight vector doesn't change is counted as $n_{\boldsymbol{w}_i}$. Then the predicted click probability of a broad keyword $k'$ can be computed as $sigmoid\left(\sum_{i=1}^{m} n_{\boldsymbol{w}_i} sigmoid(\boldsymbol{w}_i \cdot \boldsymbol{x})\right)$.

In order to effectively capture the fluid SSA environment, Gupta et al. (2009) developed an amnesiac averaged perceptron algorithm by incorporating the exponentially weighted moving average into the MMVP and exploiting implicit feedback (i.e., advertising click-through logs) to identify high-quality broad matches for a given keyword. Specifically, besides $\boldsymbol{w}$, another moving average weight vector $\boldsymbol{w}_{avg}$ is initialized as 0; for each $(\boldsymbol{x} = f(k \to k'), c)$, after updating $\boldsymbol{w}$, $\boldsymbol{w}_{avg}$ is updated as $\boldsymbol{w}_{avg} := (1 - \zeta)\boldsymbol{w}_{avg} + \zeta\boldsymbol{w}$, where $\zeta \in (0,1]$ is the amnesia rate indicating that the weight vectors observed in the past are less influential than the most recent one. Finally, the moving average weight vector is used to conduct the click probability prediction. The MMVP-based matching method can quickly adjust to rapidly-changing distributions of keywords, advertisements and user behaviors.

**(3) Distributed language model (DLM)**. DLM learns word representations in a low-dimensional continuous vector space using a surrounding context of a word in a sentence, where in the resulting embedding space, semantically similar words are close to each other (Mikolov et al., 2013). In SSA, DLM can be used to learn ad and query representations in a low-dimensional space and solve the query-ad matching problem (Grbovic et al., 2016). Given a search session set $\hat{S}$, $s = (a_1, \ldots, a_N) \in \hat{S}$ is an uninterrupted sequence of user's actions comprising queries, ad clicks, and link clicks. The search embedding learns a D-dimensional real-valued representation $\boldsymbol{v}_{a_n} \in \mathbb{R}^D$ for each unique action $a_n$ by maximizing the following objective function

$$\hat{\mathcal{L}} = \sum_{s \in \hat{S}} \sum_{a_n \in s} \sum_{-\xi \leq i \leq \xi, i \neq 0} log\, p(a_{n+i}|a_n)$$
$$= \sum_{s \in \hat{S}} \sum_{a_n \in s} \sum_{-\xi \leq i \leq \xi, i \neq 0} log\, \frac{\exp(v_{a_n}^T v'_{a_{n+i}})}{\sum_{a=1}^{|\Phi|} \exp(v_{a_n}^T v'_{a_n})}, \quad (36)$$

where $\boldsymbol{v}_a$ and $\boldsymbol{v}'_a$ are the input and output vector representations of user's actions, respectively, $\xi$ is the length of the relevant context for action sequences, and $\Phi$ is a vocabulary set for unique actions in the dataset consisting of queries, ads, and links.

The optimization can be done via stochastic gradient ascent. Grbovic et al. (2016) presented a DLM-based matching method through semantic embeddings of search queries and advertisements. This method delineates the temporal context of action sequences, where actions with similar contexts will have similar representations, and reduces the complex broad match problem to a trivial K-nearest-neighbor search between queries and ads in the joint embedding space. The DLM-based method can gain a good performance in terms of relevance, coverage and incremental revenue.

**(4) Graph model (GM)**. Broad match graph is a weighted bipartite graph between a set of advertisers and a set of keywords. Let $n_K$ denote the total number of keywords, $n_{Ader}$ denote the total number of advertisers, $n_{AS}$ denote the maximum number of ad-slot available, $\mathcal{E}$ denote the valuation matrix where $\varepsilon_{j,k}$ is the product of true value and quality score of advertiser $ader_j$ for keyword $k$, $B_j$ is the daily budget for $ader_j$ and $d_k$ denote the daily search demands of $k$. Given instance parameters $(n_{Ader}, n_K, n_{AS}, \mathcal{E} = (\varepsilon_{j,k}), B = (B_j), d = (d_k))$, a bipartite graph $G = (Ader, K, \mathcal{S})$ is constructed, with vertex sets (i.e., the advertiser set $Ader$ and the keyword set $K$) and the edge set (i.e., $\mathcal{S} = \{(ader_j, k): ader_j \in Ader, k \in K, \varepsilon_{j,k} > 0\}$). Each edge $(j, k) \in \mathcal{S}$ has a weight $\varepsilon_{j,k}$, each ader-node $ader_j \in Ader$ has a weight $B_j$, and each keyword-node $k \in K$ has a weight $d_k$. Then, given instance parameters $(n_{Ader}, n_K, n_{AS}, \mathcal{E}', B, d)$ where $\mathcal{E} \subset \mathcal{E}'$, $G' = (Ader, K, \mathcal{S}')$ is a broad-match graph for $G = (Ader, K, \mathcal{S})$, if $\mathcal{S} \subset \mathcal{S}'$ and $\varepsilon'_{j,k} = \varepsilon_{j,k}$ for all $(j, k) \in \mathcal{S}$.

Based on the broad match graph, Singh and Roychowdhury (2008) studied dynamics of bidding over various related keywords, and discussed two broad match scenarios and the extent of auctioneer's control on budget splitting. When the quality of broad match is good, the auctioneer (i.e., search engine) could always improve the revenue by judiciously using broad match.

**(5) Game-theoretic model (GTM)**. GTM uses game theory (i.e., mathematical models of strategic interactions among rational agents) to predict actions of either cooperative or competitive individuals (Moorthy, 1985). Amaldoss et al. (2016) applied a game-theoretic model to analyze advertisers' expected utilities in four possible cases, i.e., both advertisers use broad match ($EU_{B,B}$), advertiser 1 uses exact match but advertiser 2 uses broad match ($EU_{X,B}$), advertiser 1 uses broad match whereas advertiser 2 uses exact match ($EU_{B,X}$), and both advertisers use exact match ($EU_{X,X}$), under an assumption that there are two risk-neutral advertisers bidding for a keyword. Their analysis disclosed that a) search engine profits increase when advertisers adopt broad match; b) search engines should increase the accuracy of broad match up to the point where advertisers are willing to adopt.

Effective keyword match can improve both the relevance and monetization of SSA campaigns by controlling advertisers' reach (Gupta et al., 2009). Current research on keyword match concentrated on various themes: identifying high-quality broad match mappings (Gupta et al., 2009; Grbovic et al., 2016), allocating budget over several broad match scenarios (Singh and Roychowdhury, 2008), optimizing the relevance of keyword match (Radlinski et al., 2008) and examining the strategic role of broad match (Amaldoss et al., 2016). Nonetheless, in order to obtain an optimal solution for keyword targeting, keyword match should be addressed together with keyword selection.

### 4.3 Input Features for Keyword Targeting

Table 8a summarizes input/features used in keyword selection techniques reported in the literature. From Table 8a, we can see that historical performance is the main source for feature extraction in keyword selection, e.g., profit-to-cost (Rusmevichientong and Williamson, 2006), cost per click and clicks (Arroyo-Cañada and Gil-Lafuente, 2019; Symitsi et al., 2022). In machine learning based keyword selection, keyword combinations are taken as features (Kiritchenko and Jiline, 2008) and extracted from queries and keywords, combined with human judged labels (Zhang et al., 2014b).

Table 8a. Input/Features used for keyword selection

| Refs. | Features | | | | | | | | | | | |
|---|---|---|---|---|---|---|---|---|---|---|---|---|
| | PCR | WC | EV | BP | CPC | IMP | CTR | VPA | RS | CL | OS | RAP | CPR |
| Rusmevichientong &Williamson (2006) | √ | | | | | | | | | | | | |
| Kiritchenko &Jiline (2008) | | √ | √ | | | | | | | | | | |
| Zhang et al. (2014b) | | | | √ | √ | √ | √ | √ | √ | | | | |
| Arroyo-Cañada &Gil-Lafuente (2019) | | | | | √ | √ | √ | | | √ | √ | √ | |
| Symitsi et al. (2022) | | | | | √ | √ | √ | | | √ | | | √ |

Notes: PCR=Profit-to-Cost Ratio; WC=Words Combinations; EV=Engaged Visit (i.e., the time spent at the website multiplied by the number of pages visited); BP=Bid Price; CPC=Cost Per Click; IMP=Impression (e.g., impression per week, impression probability based on bid price, impression confidence based on competitiveness); CTR=Click-Through Rate; VPA=Value Per Ad-group; RS=Relevance Score (calculated based on the query-ad similarity, semantic similarity, taxonomy, and user query time); CL=Clicks; OS=Opportunity to See; RAP=Relevance of the Advertising Place; CPR=Cots-Per-Reservation.

Table 8b summarizes input/features used in research on keyword match. In keyword targeting, match type is often considered as an important decision factor where optimization and equilibrium analysis are conducted (Singh and Roychowdhury, 2008; Amaldoss et al., 2016). Feature selection provides considerable improvement in keyword match (Gupta et al., 2009). For keyword match optimization, features extracted from search sessions are used to learn low-dimensional continuous representations of queries and advertisements (Grbovic et al., 2016). Moreover, features such as lexical similarity, semantic similarity and revenue describe match quality between the query and candidate substitution (Radlinski et al., 2008).

Table 8b. Input/Features used for Keyword Match

| Refs. | Features ||||||||||||||
|---|---|---|---|---|---|---|---|---|---|---|---|---|---|---|
| | MF | CO | DE | SSM | CTR | KID | SQ | CL | CI | IA | BC | LSF | SSF | KO | SD | PRF |
| Gupta et al. (2009) | √ | √ | √ | √ | √ | √ | | | | | | | | | | |
| Grbovic et al. (2016) | | | | | | | √ | √ | √ | | | | | | | |
| Singh &Roychowdhury (2008) | | | | | | | | | | √ | √ | | | | | |
| Radlinski et al. (2008) | | √ | | | | | | | | | | √ | √ | √ | √ | √ |
| Amaldoss et al. (2016) | √ | | | | | | | | | | | | | | | |

Note: MF=Match related Feature (e.g., broad match mapping, the accuracy of broad match bid); CO=Co-occurrence; DE=Densified (i.e., the local structure of similarity graphs); SSM=Syntactic Similarity Measures (e.g., string edit distance, the presence of one keyword as a substring inside the other); CTR=Click-Through Rate; KID=Keyword-ID (e.g., the total number of bidded keywords for the original and broad-match keywords); SQ=Sequences of Queries; CL=Clicks (ad clicks, search link clicks); CI=Contextual Information (e.g., dwell time and skipped ads); IA=Information Asymmetry; BC=Budget Control (i.e., the extent of auctioneer's control on the budget splitting); LSF=Lexical Similarity Features (i.e., share words, word distance, cosine and trigram cosine); SSF=Semantic Similarity Features (i.e., max match score, abstract cosine and taxonomy similarity); KO=Keyword Overlap; SD=Search Demand; PRF= Potential Revenue Features (i.e., max bid, second bid).

## 4.4 Evaluation Metrics for Keyword Targeting

Evaluation metrics that we identified in the reviewed articles on keyword targeting (i.e., keyword selection and match) are presented in Table 9. As keyword decisions move deeper into the lifecycle framework from keyword generation to keyword targeting, evaluation metrics become closer to the ultimate goal of keyword decisions for advertisers, i.e., revenue/profit maximization. In particular, the revenue/profit is the most common evaluation metric in keyword targeting (either keyword selection or keyword match) (Rusmevichientong and Williamson, 2006; Singh and Roychowdhury, 2008; Radlinski et al., 2008; Gupta et al., 2009; Zhang et al., 2014b; Symitsi et al., 2022), while the most commonly used metrics in keyword generation (e.g., precision, recall and F-measure) disappear. Moreover, it appears that keyword selection and keyword match emphasize different metrics. That is, keyword selection highlights keyword/advertising performance indexes such as impressions, expected clicks, CPC, and advertising cost, while keyword match accentuates metrics such as NDCG, relevance, and coverage (Radlinski et al., 2008; Gupta et al., 2009; Grbovic et al., 2016).

Table 9. Evaluation Metrics for Keyword Targeting

| **Metrics** | **Definition** | **Research Stream** | **Refs.** |
|---|---|---|---|
| Revenue/Profit | An advertiser's or search engine's economic benefits. | Keyword selection | Rusmevichientong &Williamson (2006); Kiritchenko &Jiline (2008); Zhang et al. (2014b); Symitsi et al. (2022) |
| | | Keyword match | Gupta et al. (2009); Singh &Roychowdhury (2008); Radlinski et al. (2008) |
| AUC | The area under the ROC curve representing the degree or measure of separability. | Keyword selection | Kiritchenko &Jiline (2008) |
| | | Keyword match | Grbovic et al. (2016) |
| Impressions | The number of times that an advertisement is displayed on results pages. | Keyword selection | Zhang et al. (2014b); Symitsi et al. (2022) |
| Expected clicks | The number of times that the ads get clicked when shown for that keyword. $Expected\ clicks = Average\ CTR * Impression$. | Keyword selection | Zhang et al. (2014b) |
| Cost per click | An advertiser's pay for each click on the ads. | Keyword selection | Zhang et al. (2014b) |
| Advertising cost | The sum of all keyword costs which equals to the search engine revenue. | Keyword selection | Zhang et al. (2014b) |
| Brand awareness | The awareness about the brand related to impressions per week, cookies per week, opportunity to see and relevance. | Keyword selection | Arroyo-Cañada &Gil-Lafuente (2019) |
| Website Traffic | The traffic to the corporative website most lined with clicks, click-through rate and cost per click. | Keyword selection | Arroyo-Cañada &Gil-Lafuente (2019) |
| Rankings of proximities | The keyword set rankings of proximities to the ideal solution. | Keyword selection | Arroyo-Cañada &Gil-Lafuente (2019) |
| Risk | The performance of selected keywords in standard deviation of the popularity growth. | Keyword selection | Symitsi et al. (2022) |
| Sharpe ratio | The difference between investment returns and the risk-free return, divided by the standard deviation of investment returns. $Sharpe\ ratio = (R_p - R_f)/\sigma_p$, | Keyword selection | Symitsi et al. (2022) |

|  | where $R_p$ is the return of portfolio, $R_f$ is the risk-free rate, and $\sigma_p$ is the standard deviation of the portfolio's excess return. |  |  |
| --- | --- | --- | --- |
| Keyword number | The number of selected keywords. | Keyword selection | Symitsi et al. (2022) |
| LogLoss | Log-loss over a test dataset. $X = \{(k \rightarrow k', c)\}: LogL(X) = \sum_{(k \rightarrow k', c) \in X} log_2(p(c\|k \rightarrow k'))$, where $c$ is the click-through rate (CTR), and $k'$ is one of the replacements given by the broad match mapping for the original keyword $k$. | Keyword match | Gupta et al. (2009) |
| LogL-Lift | The difference between the model's log-likelihood and the entropy of the test set. | Keyword match | Gupta et al. (2009) |
| Relative CTR | The CTR of the subset of the test set that overlaps with the mapping. $CTR(BM) = p(c\|(k \rightarrow k') \in BM)$, where $c$ is the click-through rate (CTR), and $k'$ is one of the replacements given by the broad match mapping $BM$ for the original keyword $k$. | Keyword match | Gupta et al. (2009); Grbovic et al. (2016) |
| Coverage | The number of items where model made any prediction divides the number of total items, specifically the coverage of the mapping. $Coverage(BM) = p((k \rightarrow k') \in BM)$, where $k'$ is one of the replacements given by the broad match mapping $BM$ for the original keyword $k$. | Keyword match | Gupta et al. (2009); Grbovic et al. (2016) |
| Macro NDCG | A measure of how well the ranked scores align with the ranked editorial grades using $(2^{grade} - 1)$ as NDCG labels and position discounting of log. | Keyword match | Grbovic et al. (2016) |
| Relevance | The relevance of an advertisement and a query substitution. | Keyword match | Radlinski et al. (2008) |
| Utility (for search engine) | The expected utility $EU_{T_1, T_2}$ for advertiser 1 choosing $T_1$-type match and advertiser 2 choosing $T_2$-type match. | Keyword match | Amaldoss et al. (2016) |

## 4.5 Summary

In summary, as we were aware of the existing literature, there are few studies on keyword targeting (Li and Yang, 2022). Instead, scholars have tried to address either keyword selection or determination of

keyword match types separately. We argue that it's of necessity to address the keyword selection and keyword match problems in an integrated way, in order to help advertisers effectively reach the targeted population via SSA campaigns.

## 5. Keyword Assignment and Grouping

### 5.1 Problem Description

SSA is a structural advertising form, which is distinctly different from the flattened structure of traditional advertising (Yang et al., 2017). For advertisers who wish to promote their products or services via SSA, they need to design one or more advertising campaigns, and create one or more ad-groups for each campaign (we call this the basic SSA structure) (Chatwin, 2013), as shown in Figure 2. Given that a set of keywords is determined by the keyword targeting process (i.e., the target keyword set), an advertiser needs to assign a subset of target keywords to each campaign, and then each campaign-specific set of keywords also needs to be grouped into several subsets, one of which corresponds to an ad-group. Keyword assignment is conducted at the campaign level, and keyword grouping is made at the ad-group level. The output of this step is the keyword structure.

From an operational perspective, keyword assignment and grouping is one of the most critical keyword decisions throughout the entire life cycle of SSA campaigns (Yang et al., 2019; Whitney, 2022). In SSA, advertising campaigns with one or several ad-groups are run to fulfill promotional goals, which constitute the search advertising structure and serve as the basic units for daily advertising operations. Organizing keywords according to search advertising structures allows advertisers to better manage advertising activities (Rutz et al., 2012) and track the effectiveness of their advertising efforts (Hou, 2015). Moreover, keyword assignment and grouping helps advertisers display the ads to the right consumers (Gopal et al., 2011; Polato et al., 2021).

Formally, given that a set of keywords and corresponding match types $K^{(TGT)}$ are determined, keyword assignment and grouping can be defined as follows.

(1) Keyword assignment:

$$f^{(ASM)}: K^{(TGT)} \xrightarrow{x_{i,j}^{(ASM)}} K_1^{(ASM)}, \ldots, K_j^{(ASM)}, \ldots, K_{n_{campaign}}^{(ASM)},$$

$$x_{i,j}^{(ASM)} = \begin{cases} 1, & \text{if } k_i \text{ is assigned to the } j-th \text{ campaign} \\ 0, & \text{otherwise} \end{cases},$$

$$i \in \{1,2,\ldots,n^{(TGT)}\}, j \in \{1,2,\ldots,n_{campaign}\}, \qquad (37)$$

where $f^{(ASM)}$ is the keyword assignment function, $K_j^{(ASM)}$ is the set of keywords assigned to the $j$-th campaign, and $x_{i,j}^{(ASM)}$ is the decision variable of keyword assignment, indicating whether the $i$-th keyword is assigned to the $j$-th campaign.

(2) Keyword grouping:

$$f^{(GRP)}: K_j^{(ASM)} \xrightarrow{x_{i,j,l}^{(GRP)}} K_{j,1}^{(GRP)}, \ldots, K_{j,l}^{(GRP)}, \ldots, K_{j,n_{group}}^{(GRP)},$$

$$x_{i,j,l}^{(GRP)} = \begin{cases} 1, if\ k_i\ is\ \text{grouped into the}\ l-th\ \text{ad}-\text{group of the}\ j-th\ \text{campaign} \\ 0, otherwise \end{cases},$$

$$i \in \{1,2,\ldots,n^{(TGT)}\}, j \in \{1,2,\ldots,n_{campaign}\}, l \in \{1,2,\ldots,n_{group}\}, \tag{38}$$

where $f^{(GRP)}$ is the keyword grouping function, $K_{j,l}^{(GRP)}$ is the set of keywords grouped to the $l$-th ad-group of the $j$-th campaign, and $x_{i,j,l}^{(GRP)}$ is the decision variable of keyword grouping, indicating whether the $i$-th keyword is grouped into the $l$-th ad-group of the $j$-th campaign. Notations used in keyword assignment and grouping are presented in Table 10.

Table 10. Notations in keyword assignment and grouping

| Terms | Definition |
|---|---|
| $d_i$ | The total number of search demands of the $i$-th keyword in a search market |
| $c_{il}$ | The click-through rate (CTR) of the $i$-th keyword in the $l$-th ad-group |
| $p_{il}$ | The cost-per-click (CPC) of the $i$-th keyword in the $l$-th ad-group |
| $x_{il}^{(GRP)}$ | The 0-1 binary decision variable indicating whether the $i$-th keyword is assigned to the $l$-th ad-group or not. |
| $B_l$ | The advertising budget available to the $l$-th ad-group |
| $\varphi_l$ | An acceptable probability range of the $l$-th ad-group |
| $\lambda_i$ | The value-per-sale of the $i$-th keyword |
| $\rho_{il}$ | The conversion rate of the $i$-th keyword in the $l$-th ad-group |
| $\psi$ | The risk-tolerance of an advertiser |
| $\boldsymbol{w}$ | The normal vector to the hyperplane |
| $\tau$ | The intercept of the hyperplane function |
| $\boldsymbol{x}_i$ | The $i$-th training sample in the form of a multiple dimensional real vector |
| $n_{keyword}$ | The maximum number of words per keyword |
| $n_{embedding}$ | The size of the embedding |
| $f_{\boldsymbol{\theta}}$ | A multi-layer perceptron with a single hidden layer parametrized by $\boldsymbol{\theta}$ |
| $f_{\boldsymbol{\theta'}}$ | A multiple layers feed forward network with a linear activation function in the output layer parametrized by $\boldsymbol{\theta'}$ |
| $\boldsymbol{\theta}, \boldsymbol{\theta'}$ | The network parameters |

## 5.2 Keyword Assignment and Grouping

Although how to effectively organize keywords following search advertising structures (i.e., keyword assignment and grouping) is a critical operational-level issue, as far as we are aware, in the literature, less attention has been put to keyword assignment and grouping, except for recent works on keyword grouping (Li & Yang, 2020) and keyword categorization (Krasňanská et al., 2021; Polato et al., 2021). Techniques used for keyword assignment and grouping in the literature are summarized in Table 11.

Table 11. Techniques for Keyword Assignment and Grouping

| Category | Approach | Refs. |
| --- | --- | --- |
| Optimization | Chance constrained programming | Li &Yang (2020) |
| Machine learning | Linear support vector machine | Krasňanská et al. (2021) |
| | DeepSets model | Polato et al. (2021) |

**(1) Chance constrained programming (CCP)**. CCP is a technique to solve optimization problems under various uncertainties, which formulates an optimization problem ensuring that the probability of meeting a certain constraint is above a certain level (Charnes & Cooper, 1959). Let $d_i$ denote the total number of search demands of the $i$-th keyword in a search market. Let $c_{il}$ and $p_{il}$ denote the click-through rate (CTR) and cost-per-click (CPC) of the $i$-th keyword in the $l$-th ad-group, respectively. Given an advertising campaign with $m$ ad-groups and a set of keywords (i.e., $n$), let $x_{il}^{(GRP)}$ denote the 0-1 binary decision variable indicating whether the $i$-th keyword is assigned to the $l$-th ad-group or not. Then the budget constraint in keyword assignment and grouping can be formulated with chance constrained programming as

$$p\{\sum_{i=1}^{n} x_{il}^{(GRP)} d_i c_{il} p_{il} \leq B_l\} \geq \phi_l, \qquad (39)$$

where $B_l$ is the advertising budget available to the $l$-th ad-group and $\phi_l$ is an acceptable probability range indicating that the probability that the cost of the $l$-th ad-group is less than the allocated budget, is greater than or equal to a certain level.

Considering that SSA environments are essentially uncertain, Li & Yang (2020) formulated keyword grouping as a stochastic programming problem with click-through rate and conversion rate as random variables, taking into account budget constraints and advertiser's risk-tolerance as follows.

$$\max E\left[\sum_{l=1}^{m}\sum_{i=1}^{n} x_{il}^{(GRP)} d_i c_{il} (\rho_{il}\lambda_i - p_{il})\right]$$
$$s.t. \quad p\{\sum_{i=1}^{n} x_{il}^{(GRP)} d_i c_{il} p_{il} \leq B_l\} \geq \phi_l$$
$$\text{Var}\left(\sum_{l=1}^{m}\sum_{i=1}^{n} x_{il}^{(GRP)} d_i c_{il} (\rho_{il}\lambda_i - p_{il})\right) / \sum_{l=1}^{m} B_l \leq \psi$$

$\sum_{l=1}^{m} x_{il}^{(GRP)} \leq 1, x_{il}^{(GRP)} = \{0,1\}, d_i \geq 0, \lambda_i \geq 0, p_{il} \geq 0,$

$0 \leq c_{il} \leq 1, 0 \leq \rho_{il} \leq 1,$ (40)

where $\rho_{il}$ is the conversion rate of the $i$-th keyword in the $l$-th ad-group; $\lambda_i$ is the value-per-sale of the $i$-th keyword; and $\psi$ is the risk-tolerance of an advertiser. Moreover, they developed a branch-and-bound algorithm to solve their model. Their experiments illustrated that, a) the proposed method could approximately approach the optimal level; b) keyword grouping leads to a significant improvement in the profit for search advertisers with a large number of keywords.

**(2) Linear support vector machine (LSVM)**. LSVM creates the hyperplane to segregate $n$-dimensional space into classes (Abe, 2005). Given a training dataset of $n$ points $(x_i, y_i), i = 1, ..., n$, where $x_i$ is a multiple dimensional real vector, and $y_i \in \{-1,1\}$ is a binary target value indicating the class that $x_i$ belongs, the maximum-margin hyperplane that divides the set of points $x_i$ is given as

$min\|w\|$

$s.t., y_i(w^T x_i - \tau) \geq 1, i = 1, ..., n,$ (41)

where $w$ is a normal vector to the hyperplane and $\tau$ is the intercept of the hyperplane function.

Krasňanská et al. (2021) applied a one-against-one LSVM-based method to classify keywords into multiple categories. The LSVM-based method can obtain a higher accuracy rate compared with other machine learning methods such as multinomial logistic regression and multinomial Naïve Bayes.

**(3) DeepSets model (DSM)**. DSM is a designing model for machine learning tasks whose objective functions are defined on sets that are invariant to permutations (Zaheer et al., 2017). It characterizes permutation invariant functions and provides a family of functions that has a special structure helpful to design a deep network architecture. In SSA, keywords can be regarded as a set of words. In general, when a keyword is short, the sequential order of a word in that keyword is not very important when learning a suitable keyword representation. Thus, keywords can be represented as a 2D tensor in $\mathbb{R}^{n_{keyword} \times n_{embedding}}$, where $n_{keyword}$ is the maximum number of words per keyword and $n_{embedding}$ is the size of the embedding. Given a keyword $k$, the DeepSets is given as follows.

$f_{DS}(k, \theta', \theta) = f_\theta(\sum_{k \in k} f_{\theta'}(k; \theta'); \theta),$ (42)

where $f_\theta$ is a multi-layer perceptron with a single hidden layer parametrized by $\theta$, and $f_{\theta'}$ is a multiple layers feed forward network with a linear activation function in the output layer parametrized by $\boldsymbol{\theta}'$. The network parameters can be optimized by a stochastic gradient descent approach with the goal of minimizing the cross entropy loss.

Polato et al. (2021) developed a deep learning model for multilingual keyword categorization by employing the fastText multilingual word embeddings, and designed its structure based on the DeepSets model. The DSM-based method can obtain good performance on accuracy scores and computational efficiency.

In the research stream on keyword assignment and grouping, in addition to academic efforts, research in the SSA industry has explored how to represent advertisers' business objectives through the search advertising structure. Search Engine Land (2022) suggested to create campaigns (i.e., make keyword assignment decisions) to fulfill advertisers' goals, e.g., finding consumers for the product (or service), increasing brand awareness, or driving new visitors to advertiser's website, and create ad-groups (i.e., make keyword grouping decision) connected to each campaign's goal. Keyword assignment and grouping should consider various factors including the structure of advertisers' website, products (or services) offered, locations, branded keywords and non-branded keywords, different bidding options, devices, consumer intents and keyword match types (Whitney, 2022; One PPC, 2022; Zirnheld, 2020). In the meanwhile, as the budget is set at the campaign level, it is effective to conduct keyword assignment with consideration of campaign-specific budget constraints (Whitney, 2022). For keyword grouping decisions under each campaign, it is suggested to construct more granular and specific ad-groups (Hill, 2018), taking into account ad-copies (Cherepakhin, 2021). However, methods used in the industry research are not theoretically rigorous, and lack necessary details and experimental evaluations to prove the effectiveness of the methods.

Note that, in the field of information retrieval, there are two research streams related to keywords grouping, namely keyword clustering (e.g., Regelson and Fain, 2006; Ortiz-Cordova and Jansen, 2012) and query clustering (e.g., Broder, 2002; Jansen et al., 2008; Yi and Maghoul, 2009), which are beyond the scope of our review.

**5.3 Input Features used for Keyword Assignment and Grouping**

Table 12 summarizes input/features used in techniques for keyword assignment and grouping, as reported in the literature. Together with keyword performance parameters such as click-through rate, the SSA structure matters in keyword decisions (Li and Yang, 2020). In keyword categorization, statistical characteristics of keyword like TF-IDF has often been listed as input features (Krasňanská et al., 2021), and internal word structures are also used as the underlying framework for keyword representation (Polato et al., 2021).

Table 12. Input/Features for Keyword Assignment and Grouping and Related Research

| Refs. | Features | | | | | | | | |
|---|---|---|---|---|---|---|---|---|---|
| | AS | SD | CTR | CPC | CVR | VPS | RT | IRF | IWS |
| Li &Yang (2020) | √ | √ | √ | √ | √ | √ | √ | | |
| Krasňanská et al. (2021) | | | | | | | | √ | |
| Polato et al. (2021) | | | | | | | | | √ |

Note: AS=Ad Structure; SD=Search Demand; CTR=Click-Through Rate; CPC=Cost Per Click; CVR=Conversion Rate; VPS=Value Per Sale; RT=Risk Tolerance; IRF=Information Retrieval Oriented Features (e.g., TF-IDF); IWS=Internal Word Structure.

## 5.4 Evaluation Metrics for Keyword Assignment and Grouping

As the goal of keyword assignment and grouping is to fill promotional goals (e.g., maximizing the expected profit) by finding an optimal solution for segmenting a set of keywords into groups, Li and Yang (2020) took profit and return on investment as metrics. Frequently used evaluation metrics such as precision, recall, F1-score and accuracy are employed in keyword categorization. Evaluation metrics that we identified in the reviewed articles on keyword assignment and grouping and related streams are presented in Table 13.

Table 13. Evaluation Metrics for Keyword Assignment and Grouping

| Metric | Definition | Research Stream | Refs. |
|---|---|---|---|
| Profit | An advertiser's or search engine's economic benefits. | Keyword assignment and grouping | Li &Yang (2020) |
| Return on investment | The expected profit divided by the expected total cost. | Keyword assignment and grouping | Li &Yang (2020) |
| Keywords number | The number of grouping keywords. | Keyword assignment and grouping | Li &Yang (2020); Krasňanská et al. (2021) |
| Precision | The ratio of correctly predicted positive observations to the total predicted positive observation.<br>$Precision = True\ positive/$<br>$(True\ positive + False\ positive)$. | Keyword categorization | Krasňanská et al. (2021) |
| Recall | The ratio of correctly predicted positive observations to the all observations in actual class. $Recall = True\ positive/(True\ positve + False\ negaitve)$. | Keyword categorization | Krasňanská et al. (2021) |
| F1-score | The weighted average of Precision and Recall. $F1 = 2*Precision*Recall/(Precision + Recall)$. | Keyword categorization | Krasňanská et al. (2021) |
| Accuracy | The ratio of correctly predicted observations to the total observations. $Accuracy = (True\ postitive + True\ negative)/(True\ postitive + True\ negative + False\ postitive + False\ negative)$. | Keyword categorization | Polato et al. (2021) |
| Time complexity | The amount of time taken by an algorithm. | Keyword categorization | Polato et al. (2021) |

## 5.5 Summary

These prior works on keyword assignment and grouping problem could provide additional insights for advertisers in SSA. However, few are designed for the keyword assignment and grouping optimization problem following the search advertising structure, which is one of the critical research directions in the field of keyword decisions.

## 6. Keyword Adjustment

### 6.1 Problem Description

Search engines have to serve both organic and sponsored search results with low response latency in order to support better user experiences (Bai and Cambazoglu, 2019). Additionally, owing mainly to the ever-changing nature of the bidding processes, and search users' and advertisers' behaviors, the search advertising market is extremely dynamic (Yang et al., 2015; 2022). In other words, consumer behaviors (e.g., ad clicks), characteristics of advertisement (e.g., ad positions) and competitions from other advertisers would change over time (Amaldoss et al., 2016). In such a dynamic market, advertisers need to prudently adjust their advertising strategies over time, which could be a strenuous task given the level of inherent complexity of search advertising. In particular, advertisers have to track the performance of ongoing search advertising campaigns and accordingly adjust their keyword structure in real time.

Search engine allows advertisers to actively adjust their keyword decisions. In SEA, it has been well recognized that keyword adjustment is important for advertisers to precisely display their advertisements and achieve more profit (Ye et al., 2015; George, 2019). First, keyword adjustment aims to obtain a dynamic policy maximizing the expected profit for SSA campaigns during a promotional period, which is distinctly different from one or several static keyword decisions at a series of times. Second, the number of distinct keywords used by search users is enormous in practical settings and search behaviors change over time (Bartz, 2006), which exponentially increase the search space for keyword adjustment. This raises a large challenge for the computational efficiency of online keyword adjustment.

Mathematically, given the selected keyword set $K^{(TGT)}$, we define the keyword adjustment process as follows.

$$f^{(ADJ)}: K^{(TGT)} \xrightarrow{x_{i,t}^{(ADJ)}=\left(x_{i,\overline{m},t}^{(TGT)}, x_{i,j,t}^{(ASM)}, x_{i,j,l,t}^{(GRP)}\right)} K_{j,1,t}^{(ADJ)}, \ldots, K_{j,l,t}^{(ADJ)}, \ldots, K_{j,n_{group},t}^{(ADJ)},$$
$$i \in \{1,2,\ldots,n^{(TGT)}\}, j \in \{1,2,\ldots,n_{campaign}\}, l \in \{1,2,\ldots,n_{group}\}, t \in \{1,2,\ldots T\}, \quad (43)$$

where $f^{(ADJ)}$ is the keyword adjustment function, $K_{j,1,t}^{(ADJ)}$ is the adjusted keyword set in the $l$-th ad-group of the $j$-th campaign at time $t$, and $x_{i,t}^{(ADJ)}$ is a decision vector of keyword assignment, indicating the structure adjustments of the $i$-th keyword in time $t$.

## 6.2 Keyword Adjustment and Related Work

How to effectively conduct keyword adjustment in real time has become a critical problem for advertisers in SSA. However, in the literature, no study we are aware of has been reported on this issue.

Keyword spreading is a technique related to keyword adjustment, which provides indirectly valuable help for advertisers in keyword research. In more detail, keyword spreading is a technique with the goal of optimizing the expected advertising revenue, where an advertiser substitutes high-cost keywords that are likely to be intensely competitive, with a set of related long-tail keywords that are collectively of lower costs but capable of leading to an equivalent volume of traffics. Budinich et al. (2010) provided an experimental benchmark of keyword spreading, and conducted large scale simulations to pin-point that the keyword spreading technique is generally convenient and acceptable to all three parties involved in SSA.

One related research stream to keyword adjustment is dynamic bid optimization (i.e., bid adjustment), i.e., how to adjust bids over a set of keywords over time. The bid optimization problem is referred to dynamically determining bids over a subset of keywords in order to maximize advertiser's expected profit. SSA entitles advertisers to adjust their bids over keywords and rankings of their advertisements any time they want, and their payoffs can be realized in real time, which demands a dynamic equilibrium bidding strategy. As reported in an empirical research by Zhang and Feng (2011) based on two data sets containing bidding records over a sample of keywords, advertisers may engage in cyclical bid adjustments under certain conditions. Cyclical bidding patterns happen in both generalized first-price (GFP) and generalized second-price (GSP) auctions. Importantly, cyclical bid-updating behaviors emphasize the necessity of adopting a dynamic perspective when exploring equilibrium properties of bidding strategies in SSA. For more information on bid adjustment, refer to see Borgs et al. (2007), Zhou et al. (2008), Katona and Sarvary (2010), Cai et al. (2017), and Küçükaydin et al. (2019).

Another related research stream to keyword adjustment is budget adjustment, i.e., how to allocate advertising budget over time. In addition to bid adjustment, advertisers need to dynamically distribute their advertising budgets in order to avoid early ineffective clicks and save money for better advertising opportunities in the future. In SSA, there exist three levels of budget decisions throughout the entire lifecycle of advertising campaigns, namely, allocation across SSA markets, distribution over a series of time slots, and adjustment of the daily budget across keywords (Yang et al., 2012). For more information on budget adjustment, refer to see Yang et al. (2012, 2014, 2015) and Zhang et al. (2014a).

Moreover, research in the SSA industry believes that advertisers should adjust their keyword structures in real time (Search Engine Land, 2022). However, it takes time for advertisers to manage, track and adjust advertising structures to get the optimal results (Whitney, 2022). A good search advertising structure is critical for a successful SSA efforts, which helps advertisers effectively reach the right consumers (Whitney, 2022; Search Engine Land, 2022; Cherepakhin, 2021; One PPC, 2022). As a content marketing specialist at WordStream wrote, "Not having a well-structured account is like attempting to drive a car that's not properly built – accidents are bound to happen". Saravia (2020) suggested advertisers to conduct keyword research and adjust ad-structure related settings over time for optimal advertising performance. In order to equip with a more organized and manageable SSA structure, advertisers should keep fine-tuning and optimizing their campaigns (Hill, 2018).

# 7. Discussion and Future Directions

## 7.1 General Discussion

### 7.1.1 The Disciplinary Perspective

Our review collected 43 research papers on keyword decisions that have been published in 28 outlets (journals or conferences) in the fields of computer science, artificial intelligence, information retrieval, information systems, advertising, and marketing. This suggests that keyword decisions have been a hot topic covered in a great variety of high-level journals and conferences.

From the disciplinary perspective, keyword decisions are an interdisciplinary research field, which falls into computational advertising (Yang et al., 2017). Table 14 summarizes major disciplines, terminology and publication outlets for research on keyword decisions. Note that, we tell the major discipline of a study based on the research issue it addressed, authors' affiliations and publication outlets. From Table 14, we can observe the following phenomena. First, in keyword generation, most contributions are from computer science, except for a few from management science, while for keyword targeting, keyword assignment and grouping, researchers from economics and management become the dominant forces. This is probably because that keyword generation is closely related to popular topics in computer science (e.g., query generation and expansion), while keyword targeting, keyword assignment and grouping are related to market mechanisms, advertising structure and processes.

Second, it is apparent that researchers from different disciplines have different focuses on keyword decisions. Specifically, in keyword generation, computer science researchers emphasize developing and comparing methods for keyword extraction from various sources including websites and Web pages, search result snippets, advertising databases and query logs (e.g., Yih et al., 2006; Li et al., 2007; Wu and Bolivar, 2008; Lee et al., 2009; Nie et al., 2019), and those in management science give prominence to investigating keyword generation from the perspective of consumers (e.g., Scholz et al., 2019) and using indirect associations between keywords to facilitate keyword generation (e.g., Abhishek and Hosanagar, 2007; Qiao et al., 2017; Zhang and Qiao, 2018; Zhang et al., 2021); in keyword targeting, computer science researchers primarily aim to identify a set of profitable keywords through adaptive approximation (Rusmevichientong and Williamson, 2006), feature selection (Kiritchenko and Jiline, 2008) and integer programming (Zhang et al., 2014b), and improve the matching efficiency by using machine learning methods (e.g., Radlinski et al., 2008; Gupta et al., 2009; Grbovic et al., 2016), those in economics put a premium on the balance between risk and profit in keyword selection (Symitsi et al., 2022) and the economic consequence of matching mechanisms (Singh and Roychowdhury, 2008), and those in management science concentrate on the strategic role of keyword management costs (Amaldoss et al., 2016) and brand awareness and traffic generated from selected keywords (Arroyo-Cañada and Gil-Lafuente, 2019); in keyword assignment and grouping, computer science researchers employ machine learning methods to increase accuracy and efficiency of keyword categorization (Polato et al., 2021; Krasňanská et al., 2021), and those in management science take into account search advertising structure to formulate keyword grouping in a stochastic programing framework aiming to maximize the expected profit (Li and Yang, 2020).

Third, it is apparent that there are no commonly agreed definitions for related concepts identified in the literature on keyword decisions. In particular, these terms have been defined variously in different papers. For example, regarding keyword generation, some studies used keyword suggestion (e.g., Chen et al., 2008; Sarmento et al., 2009; Schwaighofer et al., 2009; Qiao et al., 2017) to represent it, some used keyword recommendation (e.g., Thomaido and Vazirgiannis, 2011; Zhang et al., 2012a), others used keyword extraction (e.g., Yih et al., 2006; Li et al., 2007; Zhou et al., 2007; Wu and Bolivar, 2008), and there are also other terms such as keyword enrichment (GM et al., 2011), term recommendation (Bartz et al., 2006), bidterm suggestion (Chang et al., 2009), and bid phrases generation (Ravi et al., 2010). Some scholars used keyword selection to indicate the process of shrinking the domain-specific keyword pool in order to target potential customers precisely (e.g., Rusmevichientong and Williamson, 2006), while others used it to represent keyword extraction from Web pages or search logs (e.g., Berlt et al., 2011). We also found that some studies used keyword suggestion to represent keyword selection (e.g., Zhang et al., 2014b; Zhang and Qiao, 2018). We believe that this phenomenon can be attributed to the interdisciplinary nature of keyword decisions.

Table 14. The Summary of Disciplines, Terminology and Publication Outlets for Keyword Decision Research

| Keyword Decision | Refs. | Major Discipline | Journal/Conference | Terminology |
|---|---|---|---|---|
| Keyword generation | Bartz et al. (2006) | Computer science | The 2nd Workshop on Sponsored Search Auctions (EC'06) | Keyword recommendation |

| | Joshi &Motwani (2006) | Computer science | The 6th IEEE International Conference on Data Mining-Workshops (ICDMW'06) | Keyword generation |
|---|---|---|---|---|
| | Yih et al. (2006) | Computer science | The 15th International Conference on World Wide Web (WWW'06) | Keyword extraction |
| | Abhishek &Hosanagar (2007) | Management science | The 9th International Conference on Electronic Commerce (ICEC'07) | Keyword generation |
| | Li et al. (2007) | Computer science | The 1st International Workshop on Data Mining and Audience Intelligence for Advertising (ADKDD'07) | Keyword extraction |
| | Zhou et al. (2007) | Computer science | Integration and Innovation Orient to E-Society | Keyword extraction |
| | Amiri et al. (2008) | Computer science | The International Conference on Information and Knowledge Engineering (IKE'08) | Keyword suggestion |
| | Chen et al. (2008) | Computer science | The 2008 International Conference on Web Search and Data Mining (WSDM'08) | Keyword suggestion |
| | Fuxman et al. (2008) | Computer science | The 17th International Conference on World Wide Web (WWW'08) | Keyword generation |
| | Wu &Bolivar (2008) | Computer science | The 17th International Conference on World Wide Web (WWW'08) | Keyword extraction |
| | Chang et al. (2009) | Computer science | The 18th International Conference on World Wide Web (WWW'09) | Bidterm Suggestion |
| | Lee et al. (2009) | Computer science | The 32nd International ACM SIGIR Conference on Research and Development in Information Retrieval (SIGIR'09) | Keyword extraction |
| | Wu et al. (2009) | Computer science | The 18th International Conference on World Wide Web (WWW'09) | Keyword generation |
| | Sarmento et al. (2009) | Computer science | The 3rd International Workshop on Data Mining and Audience Intelligence for Advertising (ADKDD'09) | Keyword suggestion |
| | Schwaighofer et al. (2009) | Computer science | The 3rd International Workshop on Data Mining and Audience Intelligence for Advertising (ADKDD'09) | Keyword suggestion |
| | Mirizzi et al. (2010) | Computer science | The 19th ACM International Conference on Information and Knowledge Management (CIKM'10) | Keyword suggestion |
| | Ravi et al. (2010) | Computer science | The 3rd ACM International Conference on Web Search and Data Mining (WSDM'10) | Bid phrases generation |

| | Welch et al. (2010) | Computer science | The 19th ACM International Conference on Information and Knowledge Management (CIKM'10) | Keyword selection |
|---|---|---|---|---|
| | Berlt et al. (2011) | Computer science | Journal of Information and Data Management | Keyword selection |
| | GM et al. (2011) | Computer science | The 4th ACM International Conference on Web Search and Data Mining (WSDM'11) | Keyword enrichment |
| | Scaiano &Inkpen (2011) | Computer science | The International Conference on Recent Advances in Natural Language Processing | Keyword selection |
| | Thomaidou &Vazirgianni s (2011) | Computer science | The 2011 International Conference on Advances in Social Networks Analysis and Mining | Keyword recommendation |
| | Zhang et al. (2012a) | Computer science | ACM Transactions on Intelligent Systems and Technology (TIST) | Keyword recommendation |
| | Jadidinejad &Mahmoudi (2014) | Computer science | Journal of Computer & Robotics | Keyword suggestion |
| | Qiao et al. (2017) | Management science | Information & Management | Keyword suggestion |
| | Zhang &Qiao (2018) | Management science | The 22nd Pacific Asia Conference on Information Systems (PACIS'18) | Keyword suggestion |
| | Nie et al. (2019) | Computer science | IEEE Intelligent Systems | Keyword generation |
| | Scholz et al. (2019) | Management science | Decision Support Systems | Keyword generation |
| | Zhou et al. (2019) | Computer science | The 27th International Conference on World Wide Web (WWW'19) | Keyword generation |
| | Zhang et al. (2021) | Management science | Electronic Commerce Research | Keyword suggestion |
| Keyword targeting | Rusmevichie ntong &Williamson (2006) | Computer science | The 7th ACM Conference on Electronic Commerce (EC'06) | Keyword selection |
| | Kiritchenko &Jiline (2008) | Computer science | The Workshop on New Challenges for Feature Selection in Data Mining and Knowledge Discovery at ECML/PKDD 2008 | Keyword optimization |
| | Zhang et al. (2014b) | Computer science | Information Processing & Management | Keyword suggestion |
| | Arroyo-Cañada and | Management science | Operational Research International Journal | Keyword selection |

|  | Gil-Lafuente (2019) |  |  |  |
|---|---|---|---|---|
|  | Symitsi et al. (2022) | Economics | European Journal of Operational Research | Keyword selection |
|  | Radlinski et al. (2008) | Computer science | The 31st Annual International ACM SIGIR Conference on Research and Development in Information Retrieval (SIGIR'08) | Keyword match |
|  | Singh and Roychowdhury (2008) | Economics | The 4th Workshop on Ad Auctions, Conference on Electronic Commerce (EC'08) | Keyword match |
|  | Gupta et al. (2009) | Computer science | The 15th ACM SIGKDD International Conference on Knowledge Discovery and Data Mining (KDD'09) | Keyword match |
|  | Amaldoss et al. (2016) | Management science | Marketing Science | Keyword match |
|  | Grbovic et al. (2016) | Computer science | The 39th International ACM SIGIR conference on Research and Development in Information Retrieval (SIGIR'16) | Keyword match |
| Keyword assignment and grouping | Li and Yang (2020) | Management science | International Journal of Electronic Commerce | Keyword assignment and grouping |
|  | Krasňanská et al. (2021) | Computer science | TEM Journal | Keyword categorization |
|  | Polato et al. (2021) | Computer science | 2021 IEEE Symposium Series on Computational Intelligence (SSCI'21) | Keyword categorization |

### 7.1.2 The Perspective of Keyword Decisions

Following the framework of keyword decisions, articles collected in our review can be categorized into four groups: domain-specific keyword pool generation (30 studies), keyword targeting (10 studies), keyword assignment and grouping (3 studies) and keyword adjustment (0 study). From the literature distribution over different keyword decisions identified in SSA, we have the following observations.

First of all, research efforts on keyword decision topics are unbalanced. More specifically, research efforts focus more on keyword generation and keyword targeting; however, there are few research efforts systematically solving keyword optimization problems related to advertising structures defined by search engines, i.e., keyword assignment and grouping and keyword adjustment. In practical campaign management, it is impossible to exaggerate the importance of the SSA structure in keyword decisions. Although the industry needs have been clear, the academic literature on keyword assignment and grouping has been sparse. Second, concerning keyword targeting, as discussed in Section 4, a dozen of prior research efforts have separately explored either keyword selection or keyword match. However, there are few studies considering both of the two issues to realize the optimal targeting for SSA campaigns. In addition, there is no research on keyword adjustment reported in the existing literature.

Regarding the less research on keyword assignment and grouping (Section 5.2) and keyword adjustment (Section 6.2), there are several possible explanations. The major reason might be the fact that SSA is distinctly different from the traditional advertising forms due to its advertising mechanism and structure. That is, there are different related keyword decisions throughout the lifecycle of SSA campaigns (Yang et al., 2019), rather than a single decision. Moreover, the SSA environment is extremely dynamic (Zhang and Feng, 2011; Yang et al., 2022) and highly uncertain (Li and Yang, 2020). The dynamic complexity inherent in SSA makes keyword decisions complicated. Furthermore, different from budget adjustment and bid adjustment that have been widely explored by researchers, keyword adjustment is a discrete optimization problem in a large feasible space raised by the scale of keyword number. Finally, keyword decisions are interdisciplinary topics which need joint efforts from different disciplines. However, as we can realize, it's not straightforward to get over disciplinary barriers.

To sum up, although plenty of research efforts have been invested in keyword decisions, there is few, if any, research taking into account practical decision components (e.g., the search advertising structure, dynamics). In this sense, we can see that there is a big gap between academic research and search advertising practice. Thus, this calls for substantial research efforts to address the practical difficulties that search advertisers meet in keyword decisions.

### 7.1.3 The Perspective of Datasets

Table 15 summarizes the datasets used to evaluate the performance of the proposed methods for various keyword decisions in the literature. We divided the data sources into four types: synthetic data generated by the authors by using computer programs, private data collected by the authors, business data provided by search engines or firms that advertise their products and services and public data shared by their owners. The first three can only be accessed by authors and those with permissions, and the fourth is available to everybody; except for the first, the other three are real-world data.

As illustrated in Table 15, 1 synthetic dataset, 19 private datasets, 10 business datasets and 1 public dataset were used in keyword generation; 3 synthetic datasets, 1 private dataset and 6 business datasets were used in keyword targeting; 1 private dataset and 3 business datasets were used in keyword assignment and grouping. First, it is obvious that almost all studies are based on datasets that are not publicly available, which may hinder the research development in this field. Second, approximately half of the prior research has access to business datasets, and the other half has to generate the necessary data for validation or collect the data through crawlers. This may also echo the gap between academic researchers and SSA practitioners. Last but not the least, we also notice that only a few studies conducted experimental evaluation on two or more datasets. Thus, it is urgent to construct benchmarking datasets for evaluating the performance of keyword decision models and algorithms.

Table 15. The Summary of Datasets for Keyword Decisions

| Keyword Decision | Refs. | Description | Data Source |
|---|---|---|---|
| Keyword generation | Bartz et al. (2006) | The advertising database and the search click logs were obtained from Yahoo Inc., including terms and URLs used by the advertisers and the searchers, respectively. | Business data |
| | Joshi &Motwani (2006) | The dataset includes 8,000 search terms picked randomly from Web pages relevant on three broad topics (travel, car-rentals, and mortgage). | Private data |
| | Yih et al. (2006) | The dataset includes 1,109 web pages randomly crawled from MSN Search results by the authors. | Private data |
| | Abhishek &Hosanagar (2007) | The corpus consists of 96 documents crawled from websites of 3 spas and 1 dental clinic by the authors. The initial dictionary was created by taking top 15 words from each page, out of which 1,087 were distinct. | Private data |
| | Li et al. (2007) | The dataset includes 2,200 TV broadcast transcripts crawled from CNN Live Today by the authors. | Private data |
| | Zhou et al. (2007) | The dataset includes 300 Web pages crawled from three Chinese websites (i.e., Sohu, Sina and 163), and 300 advertisements crawled from Google's sponsored lists by the authors. | Private data |
| | Amiri et al. (2008) | The INEX Wikipedia collection includes more than 658,000 documents, 267,625,000 terms. https://www.mpi-inf.mpg.de/departments/databases-and-information-systems/software/inex/ | Public data |
| | Chen et al. (2008) | The dataset includes 1,306,586 web pages crawled from the 150,446 ODP (Open Directory Project) categories by the authors. | Private data |
| | Fuxman et al. (2008) | A snapshot of the query logs obtained from a major search engine, including 41 million queries, 55 million URLs, and 93 million edges. | Business data |
| | Wu &Bolivar (2008) | The dataset includes 800 Web pages randomly crawled from a large pool of eBay partner websites by the authors. | Private data |
| | Chang et al. (2009) | The dataset includes 200 ads from Yahoo's sponsored search ad database, and each ad had fewer than 50 bidterms. | Business data |
| | Lee et al. (2009) | The dataset includes 10 popular drama shows and the first 5 episode scripts per show were selected from various sources (50 scripts, 3,404 scenes total) by the authors. | Private data |
| | Wu et al. (2009) | The dataset includes 100 seed category names widely spread over different topics from eBay and Amazon. For each characteristic document generation, top 400 Google search-hits were crawled by the authors. | Private data |
| | Sarmento et al. (2009) | The dataset includes a set of 84,180 advertisements and 122,099 unique keywords, compiled over a period of about 5 years. | Business data |
| | Schwaighofer et al. (2009) | The first dataset includes 10,000 advertisements and 100 keywords sampled from a randomly generated model.<br>The second dataset is derived from a corpus of almost 6 million advertisements and 19 million distinct keywords. | Synthetic data, Business data |
| | Mirizzi et al. (2010) | The dataset is a domain-specific subgraph of Dbpedia crawled starting from a set of seed nodes. | Private data |
| | Ravi et al. (2010) | The dataset includes a set of ads with advertiser-specified bid phrases from the Yahoo! Ad corpus, and each ads is associated with a landing URL. | Business data |
| | Welch et al. (2010) | The dataset includes a range of videos including 12 films, 3 clips from news and educational content, and 5 amateur clips crawled from YouTube by the authors. | Private data |
| | Berlt et al. (2011) | The dataset includes 300 Web pages extracted from a Brazilian newspaper. | Private data |
| | GM et al. (2011) | The dataset includes 95,104 websites sampled from contextual advertising system of Yahoo! and all Web page requests from these websites for a duration of 90 days, 220 million pages accounting for 21 billion impressions. | Business data |
| | Scaiano &Inkpen (2011) | The dataset records the performance of existing campaigns from an industrial partner. | Business data |
| | Thomaidou &Vazirgiannis (2011) | The dataset includes landing pages taken from eight different thematic areas promoting several products and services. | Private data |
| | Zhang et al. (2012a) | The Pagelinks dataset was crawled by the authors in September 2009. It includes 81.83 million triples and each triple represents a | Private data |

|  |  | relation where the first entity has a page link to the second entity, resulting in an initial graph with 81.83 million directed edges and 9.54 million entities. |  |
| --- | --- | --- | --- |
|  | Jadidinejad &Mahmoudi (2014) | The dataset leveraged 2008-07-14 offline XML version of Wikipedia, including more than 7 million articles crawled by the authors. | Private data |
|  | Qiao et al. (2017) | The dataset includes approximately 8,500 query logs on October 2014 crawled from Baidu Tuiguang by the authors. | Private data |
|  | Zhang &Qiao (2018) | Query logs were collected through Google Keyword Suggestion Tool by the authors, including query keywords and query volumes for 20 seed keywords. | Private data |
|  | Nie et al. (2019) | The dataset was crawled from Wikipedia by the authors, including 879 articles: 130, 104, 317 and 328 articles for four seed keywords, respectively. | Private data |
|  | Scholz et al. (2019) | The dataset records the performance of SSA campaigns for two large-scale online stores, which was provided by a company that runs more than 100 online stores worldwide and has online sales of just under Euro 7 billion a year. | Business data |
|  | Zhou et al. (2019) | The dataset includes 40 million query logs provided by Sogou.com, and each sample consists of a <ad keyword, user query> pair. | Business data |
|  | Zhang et al. (2021) | The corpus of Zhihu was crawled by the authors and the corpus for calculating the Web-based kernel function and semantic drift levels was collected from top 50 search results of Baidu. | Private data |
| Keyword targeting | Rusmevichientong &Williamson (2006) | The dataset includes 100 randomly generated problem instances. In the first experiment, each problem instance has 8,000 keywords, 200 time periods, $400 budget per period, and the number of search queries in each time period has a Poisson distribution with a mean of 40,000. In the second experiment, each problem instance has 50,000 keywords, 200 time periods, $1,000 budget per period, and the number of search queries in each time period has a Poisson distribution with a mean of 150,000. | Synthetic data |
|  | Kiritchenko &Jiline (2008) | The dataset includes 3-month records of an SME's advertising campaign on Google provided by the Epiphan Systems Inc. The company operates in a video signal processing business. In the reported period, it advertised on 388 unique keywords ranging from single words to 5-word phrases. The dataset is constructed from the company's weblogs and contains all users' queries resulting in paid clicks along with the label on users' activities. | Business data |
|  | Zhang et al. (2014b) | The dataset was obtained from Microsoft Research, including search logs and auction logs collected during one month (April | Business data |
|  |  | 2011) and the advertising database containing 31 million ad groups. |  |
|  | Arroyo-Cañada &Gil-Lafuente (2019) | The advertising performance data (e.g., impressions per week, clicks per week) obtained by the advertising planner tool of Google, and twenty-four different alternative keyword sets provided by a stock exchange brokerage service company. | Business data |
|  | Symitsi et al. (2022) | The set of relevant keywords and the relevant metrics was drawn from Google Ad Words on September 11, 2015, and the Search Volume Index (SVI) time series data were drawn from Google Trends by the authors. | Private data |
|  | Gupta et al. (2009) | The dataset was derived from two months of logs collected from the Microsoft contextual advertising system containing millions of advertising impressions based on millions of bidded keywords. | Business data |
|  | Radlinski et al. (2008) | The dataset is provided by Yahoo! Research, including 10 million most frequent queries issued to the Web search engine over a one-week period. | Business data |
|  | Singh &Roychowdhury (2008) | The authors illustrated the research scenarios via calculating examples (e.g., four advertisers and two keywords with different revenue). | Synthetic data |
|  | Amaldoss et al. (2016) | The authors illustrated the research scenarios via calculating examples (e.g., a search advertising market with two risk-neutral advertisers and one keyword). | Synthetic data |
|  | Grbovic et al. (2016) | The dataset includes more than 126.2 million unique queries, 42.9 million unique ads, and 131.7 million unique links, comprising over 9.1 billion search sessions on Yahoo Search. | Business data |
| Keyword assignment and grouping | Li &Yang (2020) | The first dataset was obtained from advertising campaigns run by an e-commerce firm that promoted celebration commodity on Amazon from June 2016 to March 2017.<br>The second data set records advertising campaigns on Google adwords provided by a large firm selling sportswear, from January 2016 to September 2016. | Business data |
|  | Krasňanská et al. (2021) | The dataset includes 112,117 keywords collected by the authors using online marketing tools (e.g., Google Search Console, Google Analytics, Marketing Miner, Ahrefs, Google Ads, Collabim, various whisperers like Google Keyword Planner or Ubersuggest's free keyword tool), the website of the customer for whom the analysis was performed, competitor websites, discussion forums, social networks, and several other resources that focus on the field of jewellery. | Private data |
|  | Polato et al. (2021) | The dataset includes users' generated keywords provided by a Spanish company in 17 different languages that have been semi-automatically categorized over the Google Product Taxonomy. | Business data |

**7.2 Future Directions**

**7.2.1 Keyword adjustment**

Most of prior research focuses on static keyword decisions without consideration of the real-time keyword performance in SSA markets. Keyword adjustment should take into account changes over time in consumer behaviors (e.g., ad clicks), characteristics of advertisement (e.g., ad positions) and competitions from other advertisers (Amaldoss et al., 2016). First of all, it calls for a systematic investigation on dynamic strategies for various keyword decisions throughout the entire lifecycle of SSA campaigns, as discussed in this survey, with consideration of the interdependence between these keyword decisions.

Second, it is of necessity to explore dynamic strategies with reference to each type of keyword decisions. As for keyword generation, advertisers should maintain and update the domain-specific keyword pool as the Web contents (e.g., websites and Web pages) and search users' behaviors (e.g., query logs and clicks) have been growing and evolving over time (Nie et al., 2019). Regarding keyword targeting, prior studies (e.g., Rutz et al., 2012) showed that it's more profitable to select general and brand keywords with broad match in the initial stage, and select specific keywords with exact match in the later stage, based on the fact that consumers would change their intentions from the exploring stage to the final purchase stage in the conversion funnel (Scholz et al., 2019). Such insights from empirical studies are valuable for designing strategies for keyword adjustment; however, they are temporally coarse-grained and thus can not support real-time operations. Dynamic strategies for keyword assignment and grouping need to consider structural constraints defined by search engines and handle the evolving interactions between keyword decisions at the two levels (i.e., keyword assignment and keyword grouping). In cases with multiple campaigns, it is also necessary to consider the dynamical relationships between keyword assignment decisions among campaigns and keyword grouping decisions among ad-groups.

Third, it is suggested to apply dynamic optimization techniques such as optimal control and reinforcement learning to deal with dynamics in keyword decisions. Different from budget adjustment and bid adjustment, keyword adjustment is a discrete optimization problem in a large feasible space characterized by the scale of keyword number. Hence, discrete optimal control can be used to adjust advertisers' keyword decisions during the entire lifecycle of SSA campaigns in a real-time manner, maximizing the expected profit over a certain time horizon.

**7.2.2 Keyword assignment and grouping**

Although research in the SSA industry gave quite a few heuristic-based methods for keyword assignment and grouping (Hill, 2018; Whitney, 2022), there is no concrete evidence for their effectiveness. In this sense, it's interesting to conduct experimental evaluations to validate and compare the performance of these heuristic-based methods based on real-world datasets collected from practical SSA campaigns.

Through drawing direct analogies between SSA markets and financial markets, Dhar and Ghose (2010) stated that like financial markets, the increasing availability of information through IT and the Internet made SSA markets more efficient over time. This inspires us to treat each keyword as a risky asset and use the modern portfolio theory to assemble an asset portfolio (i.e., keyword assignment and grouping) that maximizes expected return for a given level of risk.

Optimal strategies for keyword assignment and grouping are more complicated in that they are closely related to the search advertising structure, thus need to handle multiple objectives and take into account structural constraints from search engines. It is suggested to model keyword assignment and grouping as a bi-level optimization model and apply a multi-objective deep reinforcement learning algorithm to solve the model.

**7.2.3 Deep learning technologies for keyword decisions**

In the literature, various techniques have been employed to solve keyword decision problems, such as logistic regression (e.g., Lee et al., 2009; Berlt et al., 2011), latent Dirichlet allocation (e.g., Welch et al., 2010; Qiao et al., 2017), integer optimization (e.g., Zhang et al., 2014b; Li and Yang, 2020), random walks (e.g., Fuxman et al., 2008) and game-theoretic model (e.g., Singh and Roychowdhury, 2008; Amaldoss et al., 2016), as we discussed in Sections 3-6.

Search and electronic markets prevailing nowadays provide a large amount of microstructure data for model training and adaptive learning. In recent years, the rise of deep techniques has revived interests in applying deep learning algorithms to support and facilitate advertising decisions (e.g., Cai et al., 2017; Polato et al., 2021; Yang and Zhai, 2022). By taking keyword decisions (e.g., keyword assignment and grouping) as classification problems, a variety of deep learning models can substantially improve the model performance.

Moreover, deep learning techniques are helpful to capture complex keyword correlations and complicated advertising settings for keyword decisions (Zhou et al., 2019). The attention-based neural network module (i.e., Transformer) is specifically designed to capture interactions among elements in the input (Lee et al., 2019). Such that, attention-based models, such as set transformers (Polato et al., 2021) may work well in modeling interactions among keywords to boost keyword decisions.

The performance (e.g., clicks and conversions) of a set of similar keywords may be interdependent (Amaldoss et al., 2016). Interactions between keywords can be expressed by various graphs in terms of co-occurrence relationships, semantic relationships and performance influences that can be obtained either by analyzing search advertising logs or by extracting the corpus of vocabulary dictionaries/corpus (e.g., thesaurus dictionary, Wikipedia). Graph models such as graph neural networks (GNNs) are capable of dealing with non-Euclidean graph data by representing high-order feature interactions in the graph structure (Zhai et al., 2023). Hence, GNN can be used to model complex interactions and relationships and thus improve the performance of keyword decision models.

However, it's not trivial to address the keyword decision problems in deep learning frameworks. For example, keyword generation may pursue keywords that are diversified and relevant to the seed keywords. This calls for deep learning models with high interpretability and a deep integration with domain (e.g., advertising and marketing) knowledge.

### 7.2.4 Comparison studies

In the literature on keyword decisions in the SSA context, as we discussed in previous sections, 39 metrics and 31 datasets were used in 30 articles on keyword generation; 19 metrics and 10 datasets were used in 10 articles in the research stream of keyword targeting; and 8 metrics and 4 datasets were used in 3 articles on keyword assignment and grouping. It is apparent that there is no commonly agreed evaluation metric and benchmark dataset. In the meanwhile, most of the datasets used in prior studies are not publicly available. Moreover, methods developed for keyword decisions were typically compared with a few baselines (including simple heuristic-based methods), rather than with state-of-the-art methods.

This calls for plenty of research efforts on comparison studies to evaluate various keyword decision models. It is also worthwhile to build public datasets and design a set of evaluation metrics for each type of keyword decisions in SSA.

**7.2.5 Joint optimization**

In SSA, keyword decisions at different levels form a closed-loop decision cycle, where results from high-level decisions serve as constraints/inputs for low-level decisions, and operational results at low levels create feedbacks for decisions at high levels (Yang et al., 2019). As far as we knew, most of existing efforts deal with a single keyword decision separately. It is necessary to consider joint optimization for two or more related keyword decisions in an integrated manner by considering relationships between them in the search advertising structure. Addressing keyword assignment and grouping problem in a unified model can be viewed as an example in this direction.

Another direction for joint optimization is to characterize the interface of keyword decisions with other advertising decisions in SSA, such as budget allocation and bid determination. In SSA, besides identifying relevant keywords and designing effective keyword structure, advertisers have to optimally allocate the advertising budget and determine optimal bidding prices on keywords, in order to maximize the performance of campaigns. More specifically, the outcome of budget decisions largely defines the feasible space for keyword decisions; in the meanwhile, keyword decisions are closely related to bidding decisions in that the latter can be considered as an important parameter in the former, which of the two together determine whether and when search advertisements can be displayed to potential consumers. That is, keyword decisions need to cooperate with other decision variables (e.g., budget and bid) in an integrated way. Fortunately, several researches have emerged recently in this direction, e.g., the dual adjustment of daily budget and bids (Zhang et al., 2012b; Zhang et al. 2014a), simultaneously selecting relevant keywords and putting optimal bidding prices over these keywords (Zhang et al., 2014b). Additionally, it is also an interesting perspective to elaborate cooperative keyword decision scenarios in the context of SSA and explore corresponding optimal strategies.

**8. Conclusions**

The main objective of this paper is to present a research agenda for keyword decisions in the context of SSA. We propose an overarching framework that highlights keyword related decision scenarios throughout the entire lifecycle of SSA campaigns. Based on this framework, we conduct a substantive review of the state-of-the-art literature on keyword decisions. We cover the related issues with sufficient depth by enumerating techniques for optimizing keyword decisions, input features and evaluation metrics. We further summarize the research status, identify the research gaps, and outline interesting prospects for future exploration in this area.

Keyword decisions are a distinct topic in online advertising because state-of-the-art Web science basically takes keyword as the basic unit to offer various information services. In this sense, without loss of generality, insights and solutions yielded from keyword decisions in SSA can also shed light on other online advertising forms such as social media advertising.